\documentclass[%
amsmath,amssymb,
aps,authoryear
]{aastex62}
\usepackage{graphicx}
\usepackage{bm}
\usepackage{color}
\usepackage{textcomp}
\usepackage{gensymb}
\usepackage{amsmath}
\usepackage{amssymb}

\newcommand{\rs}{R_\odot}
\newcommand{\dash}{\text{ -- }}
\newcommand{\bb}{\bm{B}}
\newcommand{\vb}{\bm{v}}

\newcommand{\psp}{\textit{PSP}}
\newcommand{\Rb}{\bm{\mathcal{R}}}
\newcommand{\ddt}[1]{\frac{\partial #1}{\partial t}}
\newcommand{\varepsilonb}{\mbox{\boldmath$\varepsilon$}}
\begin{document}
\title{Contextual Predictions for Parker Solar Probe II: Turbulence Properties and Taylor Hypothesis}
\author{Rohit Chhiber}
\email{rohitc@udel.edu}
\affiliation{Department of Physics and Astronomy, University of Delaware, Newark, DE 19716, USA}
\author{Arcadi V.~Usmanov}
\affiliation{Department of Physics and Astronomy, University of Delaware, Newark, DE 19716, USA}
\affiliation{NASA Goddard Space Flight Center, Greenbelt, MD 20771, USA} 
\author{William H.~Matthaeus}
\affiliation{Department of Physics and Astronomy, University of Delaware, Newark, DE 19716, USA}
\affiliation{Bartol Research Institute, University of Delaware, Newark, DE 19716, USA}
\author{Tulasi N.~Parashar}
\affiliation{Department of Physics and Astronomy, University of Delaware, Newark, DE 19716, USA} 
\author{Melvyn L.~Goldstein}
\affiliation{NASA Goddard Space Flight Center, Greenbelt, MD 20771, USA}
\affiliation{University of Maryland Baltimore County, Baltimore, MD 21250, USA}
\begin{abstract}
The \textit{Parker Solar Probe} (\psp)
primary mission
extends seven years and consists of 
24 orbits of the Sun with descending perihelia culminating in a closest approach of 
\(\sim 9.8~\rs\). In the course of these orbits \psp~will pass through widely varying conditions, including anticipated large variations
of turbulence properties such as energy density, correlation scales and 
cross helicities. Here we employ 
global magnetohydrodynamics simulations with self-consistent turbulence transport and heating 
\citep{usmanov2018} to preview likely conditions that will be encountered by \psp, 
by assuming suitable boundary conditions at the coronal base. The code evolves large-scale parameters -- such as 
velocity, magnetic field, and temperature -- as well as  
turbulent energy density, cross helicity, and correlation scale.
These computed quantities provide the basis for
evaluating additional useful parameters that are 
derivable from the primary model outputs.
Here we illustrate one such possibility
in which 
computed turbulence 
and large-scale parameters are used to evaluate the accuracy of the Taylor ``frozen-in'' hypothesis along the \psp~trajectory. Apart from  the immediate purpose of anticipating turbulence conditions that \psp~will encounter, as experience is gained in comparisons of observations with simulated data, this approach will be increasingly 
useful for planning and interpretation of subsequent observations.
\end{abstract}
\section{Introduction and Background}
Fundamental questions in heliospheric physics 
concern the heating of the solar corona,
acceleration of the solar wind, and the origin of suprathermal 
energetic particles. At present 
these questions are actively debated, as we anticipate 
substantial closure based on the upcoming 
pioneering observations to be made by the \textit{Parker Solar 
Probe} (\psp)~and \textit{Solar Orbiter} (\textit{SO}) missions.
\psp~will explore closest to the Sun, within 9 \(\rs\)
of the surface, and is expected to penetrate the sub-Alfv\'enic 
magnetically-dominated coronal region \citep{fox2016SSR,chhiber2019psp1}.
These landmark missions will study 
properties of the interplanetary and coronal plasmas 
in previously unexplored regions, providing information crucial to understanding structure and dynamics of these
plasmas over a wide range of spatial scales. 
Among the several types of 
novel measurements to be made by \psp~will be measurement of the mean and 
fluctuating component of plasma density, plasma 
velocity, and electromagnetic field. 
These basic measurements 
will comprise a comprehensive 
characterization of 
\textit{plasma turbulence} at scales ranging 
from larger magnetohydrodynamic (MHD) scales to subproton kinetic scales. 
This turbulence provides several ingredients 
that are potentially crucial in interplanetary dynamics \citep{matthaeus2011SSR,bruno2013LRSP}. 
The turbulent cascade of energy is expected to fuel coronal and solar wind heating,
and therefore power the distributed acceleration of the 
solar wind \citep{matthaeus1999ApJL523,verdini2010ApJ}, in addition to  direct acceleration by the ponderomotive force of turbulent pressure gradients \citep{belcher1971ApJwavepressure,alazraki1971AA}. 
Likewise, turbulence provides scattering centers 
that control spatial transport and diffusion of 
suprathermal particles such as solar energetic particles (SEPs) 
as well more energetic galactic cosmic rays \citep{jokipii1966cosmic,chhiber2017ApJS230}. 
Turbulence may also play an important role in the acceleration and transport of suprathermal particle populations \citep{jokipii1966cosmic,tessein2013ApJ} and can mediate fast, plasmoid-induced magnetic reconnection \citep{matthaeus1986PoF}. 

In this paper we focus in particular on the \psp~and anticipate measurements of turbulence that 
it is likely to make. 
To quantify the turbulence properties -- 
energy density, cross helicity, and correlation scale -- we employ a two-scale strategy that is based on a global three-dimensional (3D) MHD simulation model
\citep{usmanov2014three,usmanov2018}. This model computes large-scale ``resolved'' MHD variables -- plasma density, velocity, magnetic field, and internal energies of protons and electrons. The model also self-consistently solves turbulence transport equations for the unresolved, or subgrid-scale turbulence quantities. Further details on the method
are provided below and in the references. We note that a similar strategy was followed in a recent study \citep{chhiber2019psp1} that examined locations of critical surfaces that are anticipated along the \psp~trajectory in its various orbits. Like that earlier study, the present work is not to be viewed as a specific, detailed prediction, but rather as a \textit{context prediction}, based on likely 
conditions of solar activity and photospheric magnetic fields that are anticipated for the \psp~mission. More detailed prediction would require use of boundary conditions suitable for (i.e., closer to) the actual time of observation. The present paper also serves as a demonstration of an approach that may be valuable to inform interpretation of \psp~data, when employed with contemporaneous or updated boundary data. 

Apart from context prediction for specific turbulence parameters, 
we will also employ the combined large-scale and subgrid data
to assess the validity of the 
Taylor ``frozen-in'' 
hypothesis \citep{taylor1938ProcRSL} along the PSP trajectory, complementing previous 
analyses based on other models \cite[e.g.,][]{matthaeus1997AIPCtrajectory,howes2014ApJ,
klein2015ApJtaylor,bourouaine2018ApJ}.

In the following Section we review briefly the two-scale physical model, 
the computational framework, and in particular the turbulence transport 
formalism. In Section \ref{results} we present results for turbulence properties, first in meridional planes, 
and then along the \psp~trajectory. The 
final results subsection examines the 
validity of the Taylor hypothesis in some detail,
along the \psp~orbits. A final section summarizes the findings. 
%
%
\section{Solar Wind Model and Turbulence Transport Model}\label{sec:model}
The large-scale resolved MHD coronal and heliospheric model that we employ 
is described in detail in \cite{usmanov2014three} and \cite{usmanov2018}. 
The large-scale equations are derived from the 
underlying primitive compressible MHD equations by the process of
Reynolds-averaging \citep[e.g.,][]{mccomb1991physics}: a physical fields, e.g., $\tilde{\mathbf{a}}$, is separated into a mean and a fluctuating component: \(\tilde{\mathbf{a}} = \mathbf{a}+\mathbf{a'}\), making use of an ensemble-averaging operation where \(\mathbf{a} = \langle \tilde{\mathbf{a}} \rangle\). 
This is a two-fluid MHD code with 
a single momentum equation and separate ion and electron temperature
equations. 
The turbulence model, consistent with the  
Reynolds-averaging approach, employs eddy viscosity, turbulent magnetic diffusivity, and subgrid turbulence energy transport equations \citep{usmanov2014three,usmanov2018}. Pressure and density fluctuations are neglected.


The large-scale model equations, with emphasis on newly added terms arising due to turbulence, are:\\
\(\bullet\) continuity equation for proton density \(\rho\)\\
\(\bullet\) momentum equation for velocity \(\vb\), 
with ponderomotive term \(- \nabla \langle |\bm{B}'|^2 \rangle/8\pi\) 
and Reynolds-stress term \( \nabla \cdot \Rb\)\\
\(\bullet\) induction equation for magnetic field \(\bm{B}\) 
with turbulent induced electric field term
\(\nabla \times \mathcal{\bm{\varepsilon}}_m\sqrt{4\pi\rho}\)\\
\(\bullet\) proton pressure (energy) equation with turbulent
energy source \(f_p Q_T(\bm{r})\)\\
\(\bullet\) electron energy equation with turbulent energy source
\((1 - f_p)Q_T(\bm{r})\).\\
The terms emphasized above represent the influence of
turbulence on the mean flow: \(\Rb = \langle\rho\vb'\vb' -
\bb'\bb'/4\pi\rangle\) is the Reynolds stress tensor,
\(\mathcal{\bm{\varepsilon}}_m = \langle\vb'\times\bb'\rangle(4\pi\rho)^{-1/2}\) is the
mean turbulent electric field, and \(\langle B'^2\rangle/8\pi\) is the
fluctuating magnetic pressure, where \(\vb'\) and \(\bb'\) are the velocity and magnetic fluctuations. \(Q_T(\bm{r})\) is the turbulent heating, which is
apportioned between protons and electrons 
according to the fraction $f_p$ that 
must be determined by kinetic physics considerations.
Recent kinetic plasma simulation and theory  
provide predictions for 
$f_p$, which increases with turbulence amplitude  \citep{wu2013prl,matthaeus2016ApJ,gary2016ApJ} and also depends on the plasma \(\beta\) \citep{parashar2018ApJ,kawazura2019pnas}. Note that the turbulent heating depends on position \(\bm{r}\).

The above set of equations is solved in a frame rotating 
with the Sun, with the natural value for 
adiabatic index \(\gamma = 5/3\).
The pressure equations
include weak 
proton-electron collisional friction terms 
involving a classical Spitzer collision time scale
\(\tau_{SE}\) \citep{spitzer1965,hartle1968ApJ151}
to model the energy exchange between the
protons and electrons by Coulomb collisions
\citep[see][]{breech2009JGRA}. We neglect the electron mass in
comparison with proton mass, as well as the 
heat flux carried by protons. The electron heat flux below \(5\text{ -- }10~\rs\) is approximated by the classical collision-dominated model of \cite{spitzer1953PhRv} \citep[see also][]{chhiber2016solar}, while above \(5 \text{ -- } 10~\rs\) we adopt Hollweg's ``collisionless'' model \citep{hollweg1974JGR79,hollweg1976JGR}. See \cite{usmanov2018} for more details. 

Closure of the above system 
requires a model for unresolved turbulence.
Although the Reynolds decomposition is not formally a scale separation, we
have in mind that the stochastic components treated as
fluctuations reside mainly at the relatively small scales. Transport equations for the fluctuations may be obtained by subtracting the mean-field equations from the full MHD equations and averaging the difference  \citep[see][]{usmanov2014three}. This yields the set of equations \citep{breech2008turbulence,usmanov2014three,usmanov2018}:

\begin{equation}
\begin{aligned}
\ddt{Z^2} &+ (\vb\cdot\nabla)Z^2 + \frac{Z^2(1 - \sigma_D)}{2}\nabla\cdot\bm{u}
  + \frac{2}{\rho}\Rb\colon\nabla\bm{u} 
  + 2\varepsilonb_m\cdot(\nabla\times\bm{V}_A)  \\
  &- (\bm{V}_A\cdot\nabla)(Z^2\sigma_c) + Z^2\sigma_c(\nabla\cdot\bm{V}_A)
  = - \alpha f^+(\sigma_c)Z^3/\lambda, \label{eq:z2}  
\end{aligned}
\end{equation}
\begin{equation}
\begin{aligned}
\ddt{(Z^2\sigma_c)} &+ (\vb\cdot\nabla)(Z^2\sigma_c)
  - (\bm{V}_A\cdot\nabla)Z^2 + \frac{Z^2\sigma_c}{2}\nabla\cdot\bm{u}
  + \frac{2}{\rho}\Rb\colon\nabla\bm{V}_A   \\
  &+ 2\varepsilonb_m\cdot(\nabla\times\bm{u})
  + (1 - \sigma_D)Z^2\nabla\cdot\bm{V}_A
  = - \alpha f^-(\sigma_c)Z^3/\lambda, \label{eq:z2sig2}
\end{aligned}
\end{equation}
\begin{equation}
\ddt{\lambda} + (\vb\cdot\nabla)\lambda = \beta f^{+}(\sigma_c)Z,
\label{eq:lambda}
\end{equation}
where \(\vb\) and \(\bm{u}\) are velocities in the Sun-corotating frame and the inertial frame, respectively. The descriptors of turbulence that we
treat as dependent variables are: \(Z^2 = \langle v'^2 + b'^2\rangle\), i.e., 
twice the fluctuation energy
per unit mass where \(\bm{b}' = \bb'(4\pi\rho)^{-1/2}, \sigma_c
= 2\langle\vb'\cdot\bm{b}'\rangle/Z^2\), 
which is the normalized
cross helicity (normalized cross-correlation between velocity and magnetic
field fluctuations), 
and $\lambda$, a correlation
length perpendicular to the mean magnetic field.
Other notations are: \(\bm{V}_A = \bb(4\pi\rho)^{-1/2}\) is the mean Alfv\'en velocity, \(\sigma_D = \langle v'^2 -b'^2\rangle/Z^2\) is the normalized energy difference that we continue treating as a constant parameter (\(=-1/3\)) derived from
observations, \(\alpha\) and \(\beta\) are the K\'arm\'an-Taylor
constants \cite[see][]{matthaeus1996jpp,smith2001JGR,breech2008turbulence}, and \(f^{\pm}(\sigma_c) = (1 - \sigma_c^2)^{1/2}[(1 + \sigma_c)^{1/2} \pm (1 - \sigma_c)^{1/2}]/2\) is a function of only \(\sigma_c\) \citep{matthaeus2004grl}. The last term on the right-hand side of Equation \eqref{eq:z2} is the von K\'arm\'an 
turbulence heating rate \citep{karman1938prsl}
adapted for MHD \citep{hossain1995PhFl,wan2012JFM697,bandyopadhyay2018prx} and plasma \citep{wu2013prl}. The fluctuation energy loss due to von K\'arm\'an decay 
is balanced in a quasi-steady state by internal energy supply in the 
pressure equations, with \(Q_T = \alpha f^+(\sigma_c)Z^3/(2\lambda)\). To evaluate the Reynolds stress we assume that the turbulence is transverse to the mean field and axisymmetric about it \citep{oughton2015philtran}, so that we obtain \(\Rb/\rho = K_R(\bm{I} - \hat{\bm{B}}\hat{\bm{B}})\), where \(K_R = \langle v'^2 - b'^2\rangle/2 = \sigma_D Z^2/2\) is the residual energy, \(\bm{I}\) is the identity matrix, and \(\hat{\bm{B}}\) is a unit vector in the direction of \(\bm{B}\). For further details see \cite{usmanov2018}.

%
\section{Results \label{results}}
The simulation runs that have been 
employed for studying heliospheric structure, for comparison 
with existing spacecraft data \citep{usmanov2011solar,usmanov2012three,usmanov2014three,
chhiber2017ApJS230,chhiber2018apjl,usmanov2018}, and for context predictions \citep{chhiber2019psp1}, 
have typically been of two major types, distinguished by the
inner surface magnetic boundary condition: 
In the first type a Sun-centered 
dipole magnetic field is imposed at the inner boundary, with a specified 
tilt angle relative to the solar rotation axis.
Zero or small tilt angle is often associated with 
solar activity minimum, while larger tilt angles 
are a suitable approximation for 
the more disordered heliosphere during 
solar maximum conditions \citep{owens2013lrsp}.\footnote{\footnotesize{\psp~has been launched during solar minimum (August 2018), and solar activity is expected to rise toward the final stages of the mission \citep{fox2016SSR}.}}  
The other kind of inner magnetic boundary condition 
is one derived from suitably normalized magnetograms \citep{riley2014SoPh,usmanov2018}. The latter type may 
be construed as more realistic, but not exact, as they are 
specific to a particular Carrington rotation. 
Here we are interested in more generic conditions, 
so we will employ only the tilted dipole-type boundary conditions. 
Tilt angles of 0\degree, 10\degree, and 60\degree~will be employed in the
results illustrated here. These runs are identical in other 
parameters, and in what follows 
will be distinguished simply by referring 
to their respective tilt angles. Note that preliminary analyses of magnetogram-based simulations (not shown here) yield results similar to those presented below, with solar minimum and solar maximum magnetogram-based runs showing qualitative agreement with low and high dipole-tilt runs, respectively, as expected.

The simulation domain extends from the coronal base at $1~\rs$ to 5 au. The input parameters specified at the coronal base include: the
driving amplitude of Alfv\'en waves (30 km~s$^{-1}$), the
density ($1 \times 10^8$ particles cm$^{-3}$), the correlation scale of turbulence (\(10,500\)~km), and temperature
($1.8 \times 10^6$~K). The cross helicity in the initial state is set as \(\sigma_c = -\sigma_{c0} B_r/B_r^\text{max}\), where \(\sigma_{c0}=0.8\), \(B_r\) is the radial magnetic field, and \(B_r^\text{max}\) is the maximum absolute value of \(B_r\) on the inner boundary. The magnetic field magnitude is assigned using a source magnetic dipole on the Sun's poles (with
strength 12 G to match values observed by Ulysses). The input parameters also include the fraction of turbulent energy absorbed by protons $f_p = 0.6$. Further details on the numerical approach and initial and boundary conditions may be found in \cite{usmanov2018}.
\subsection{Turbulence Parameters in Meridional Planes}
\begin{figure}
\gridline{\fig{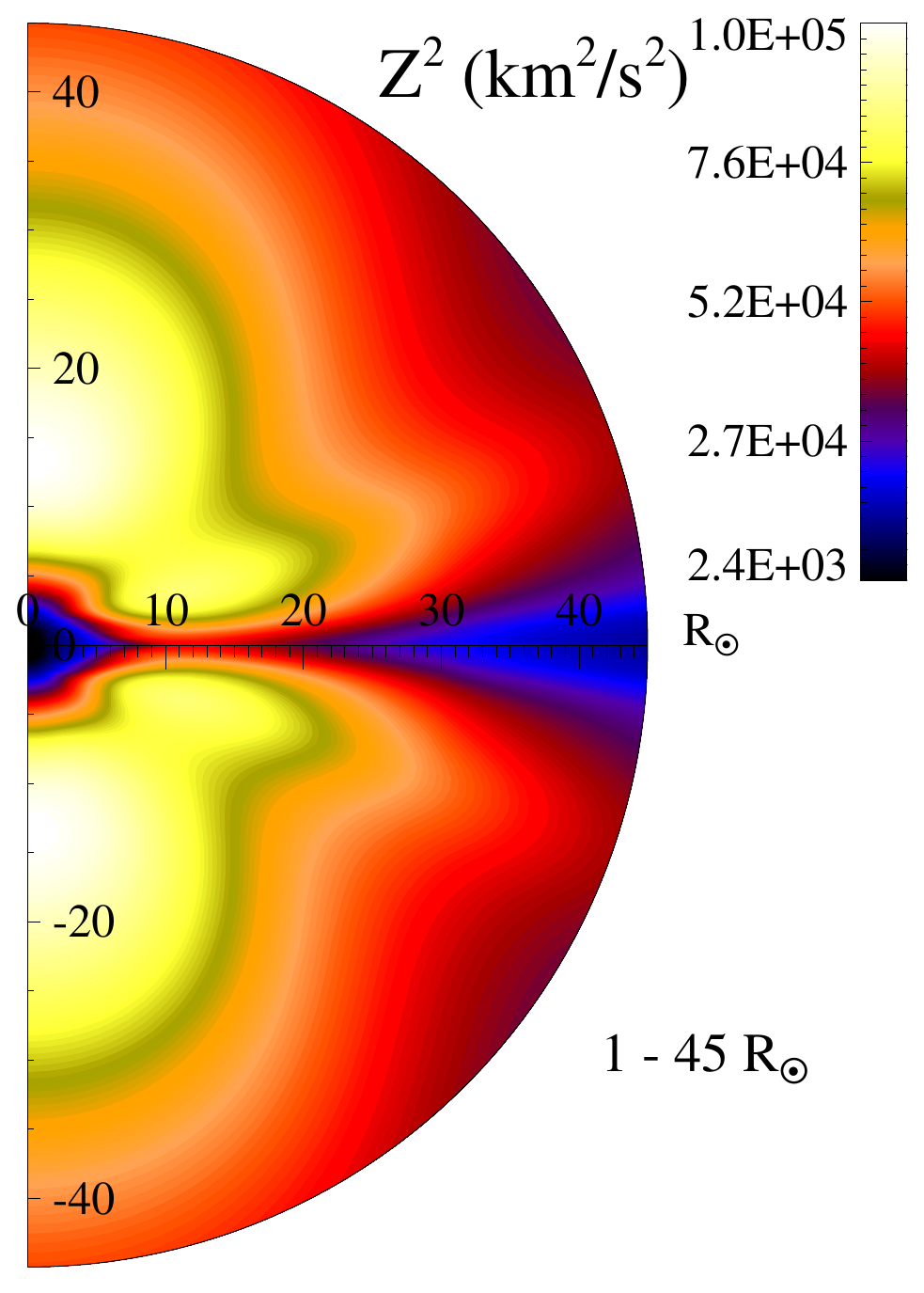}{0.25\textwidth}{}
          \fig{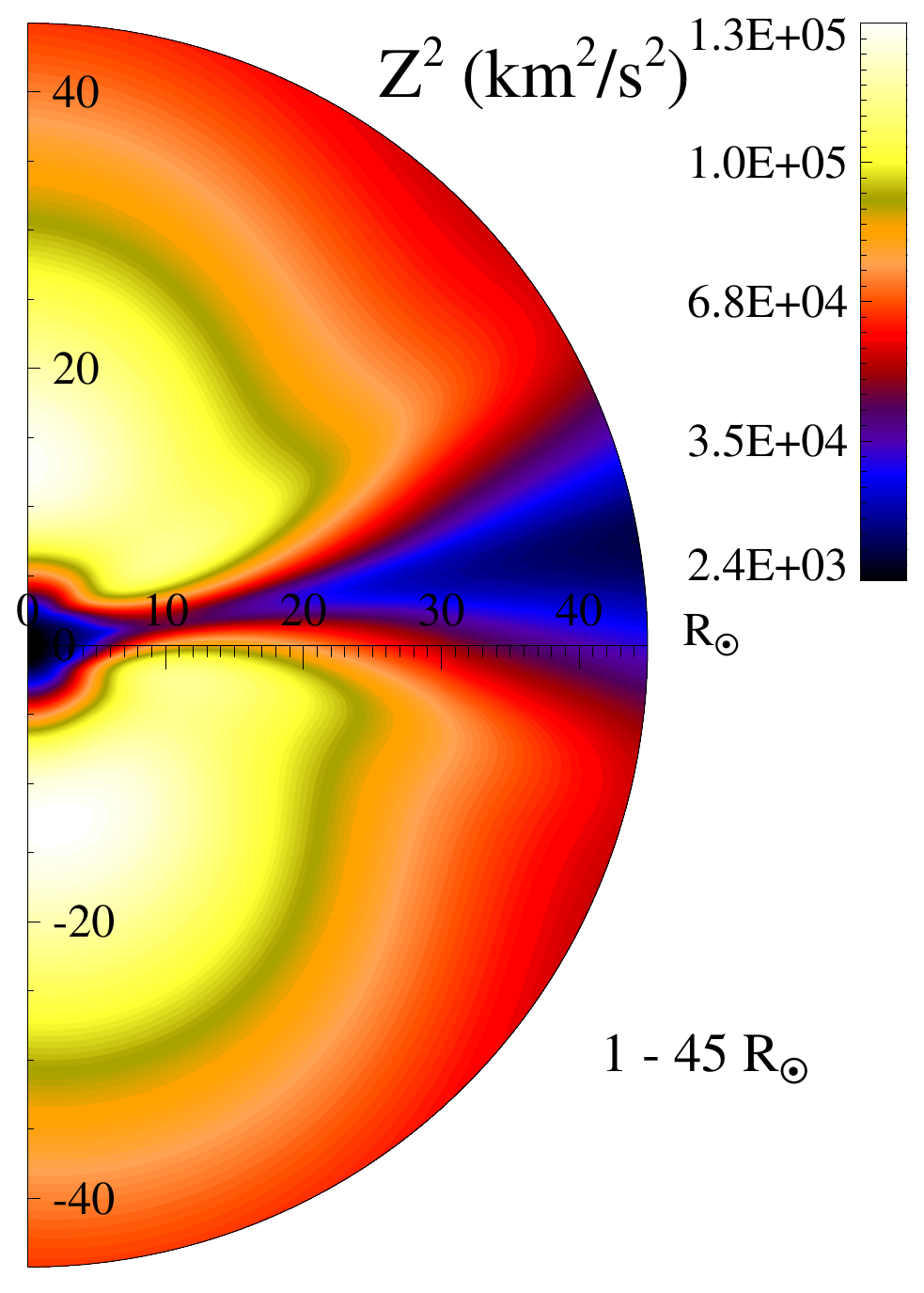}{0.25\textwidth}{}
          \fig{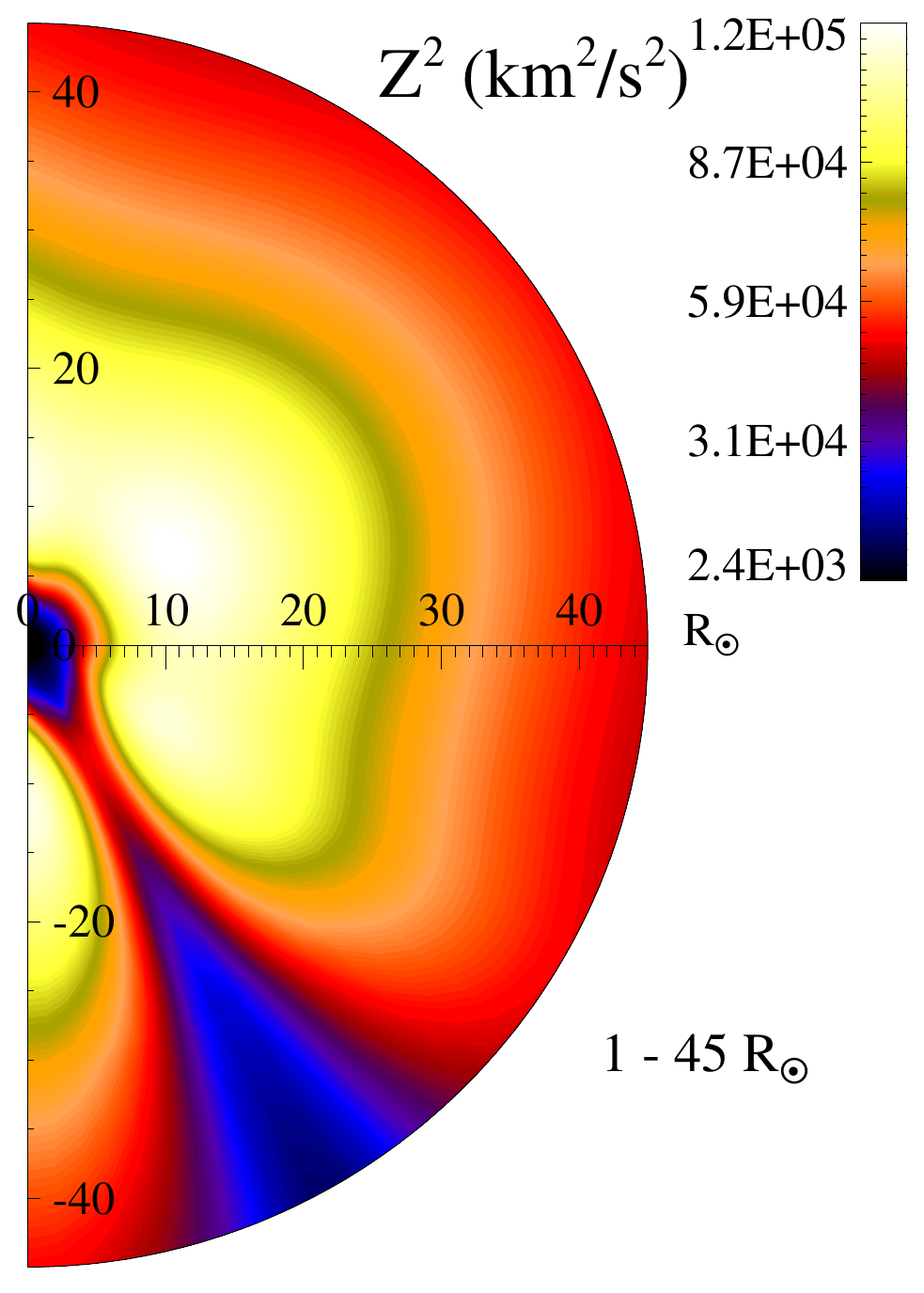}{0.25\textwidth}{}
          }
\gridline{\fig{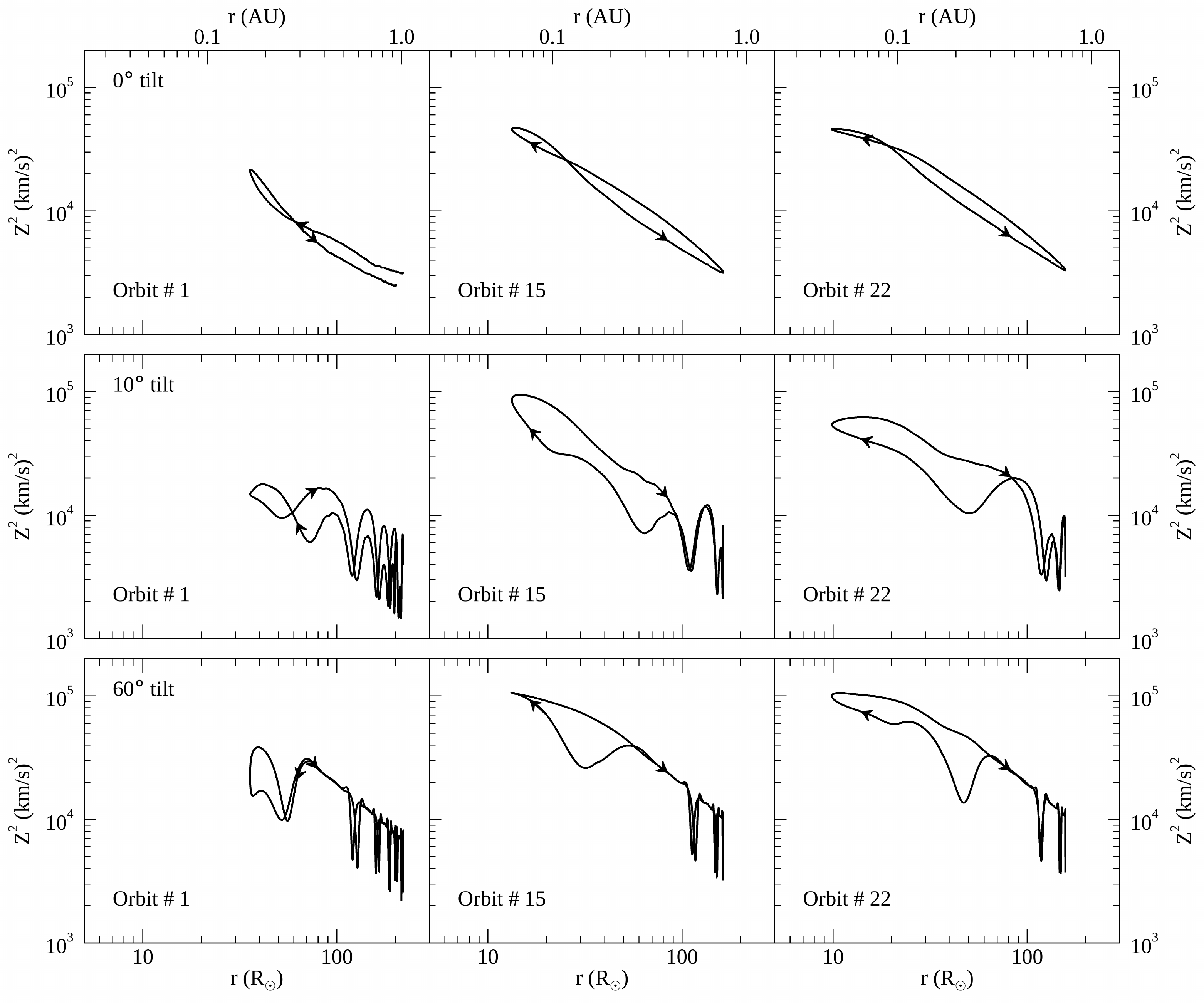}{.9\textwidth}{}}
\caption{Top panel shows turbulence energy density \(Z^2\) in a meridional plane in the region~\(1\dash 45~\rs\), from simulations with source dipoles tilted by (left) 0\degree, (middle) 10\degree, and (right) 60\degree~relative to the solar rotation axis. The next three panels show \(Z^2\) along the \psp~trajectory for selected orbits, for the three dipole tilts. Direction of arrows indicates inbound/outbound sections of orbits.}
\label{fig:z0}
\end{figure}

\begin{figure}
\gridline{\fig{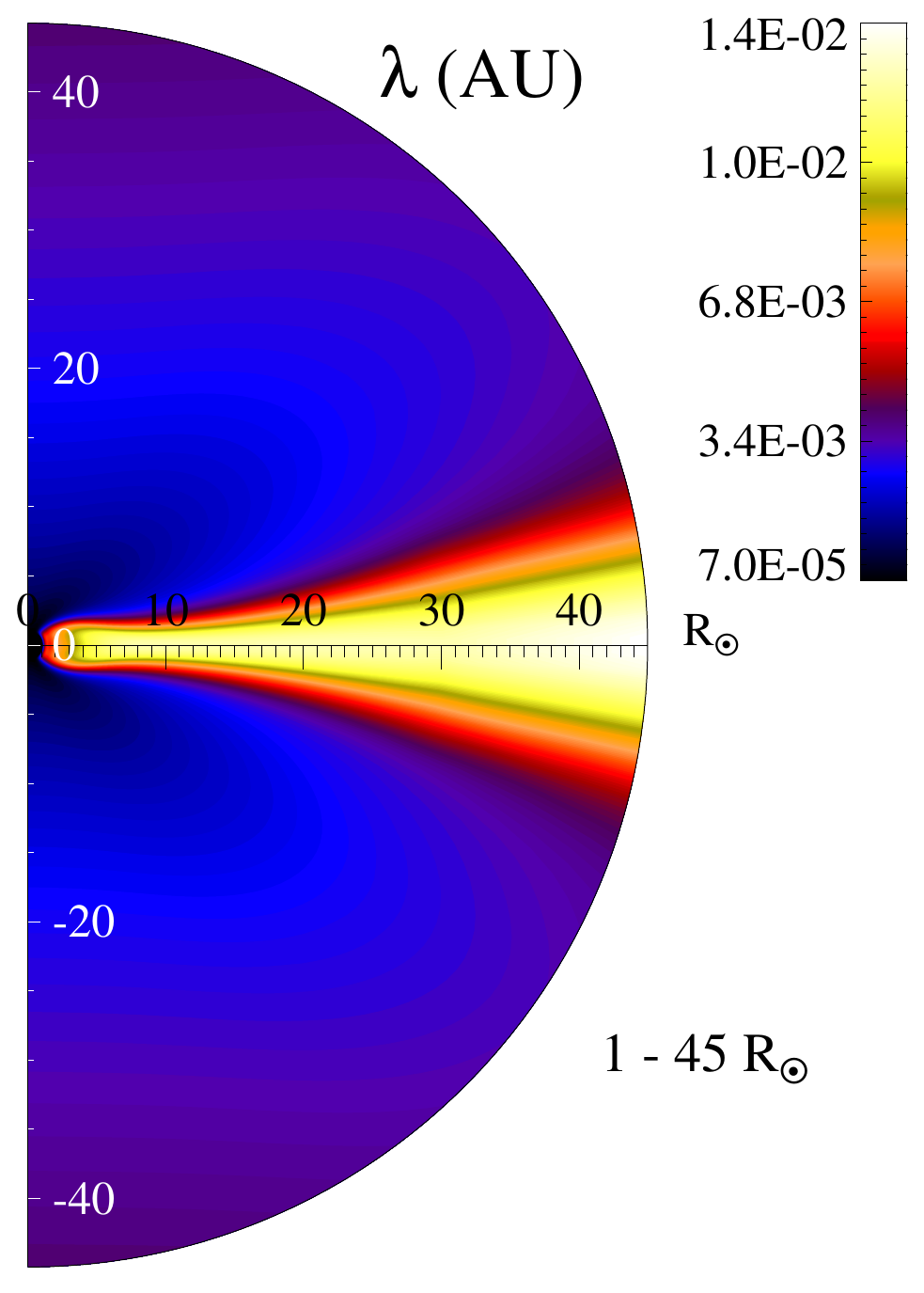}{0.25\textwidth}{}
          \fig{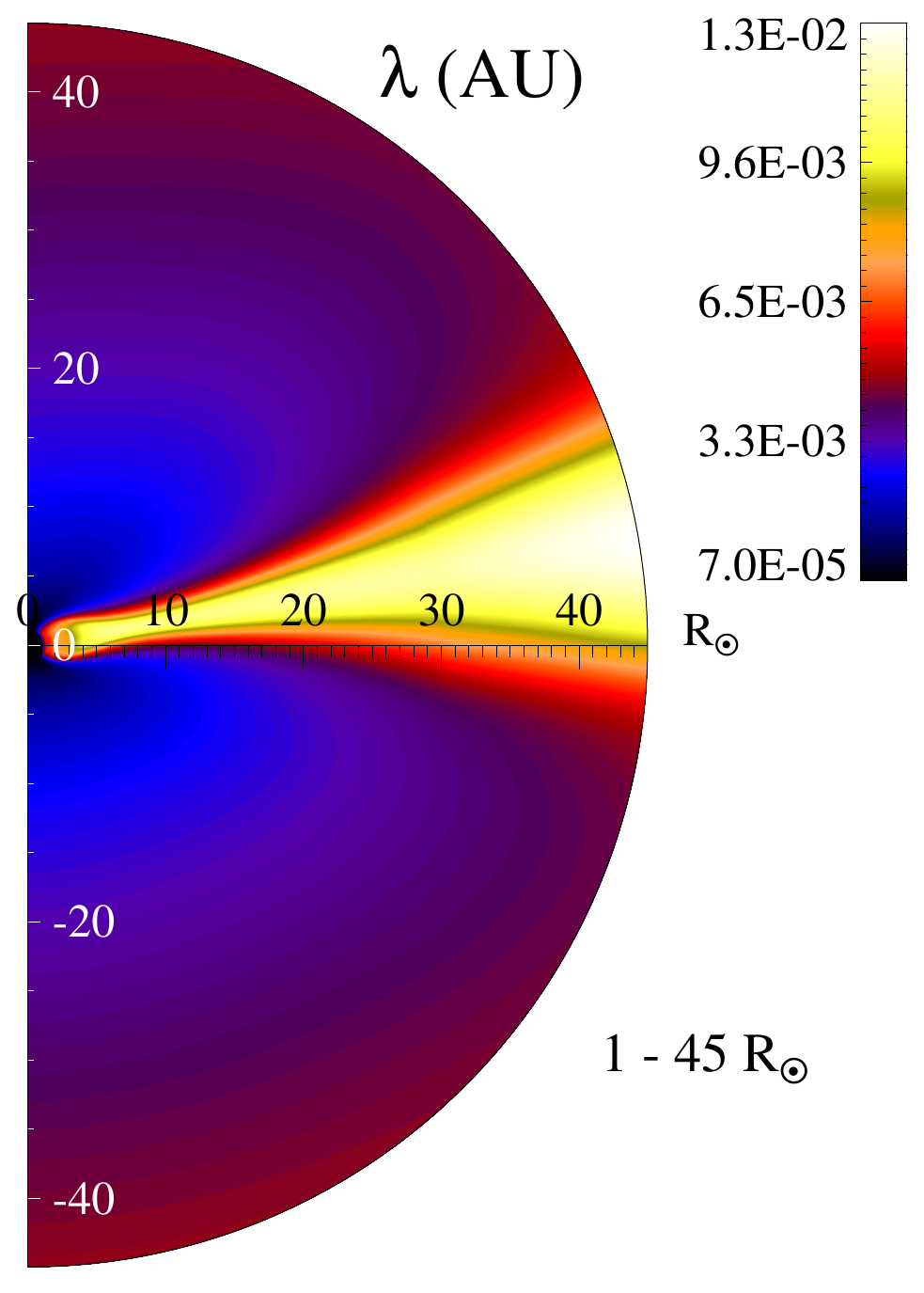}{0.25\textwidth}{}
          \fig{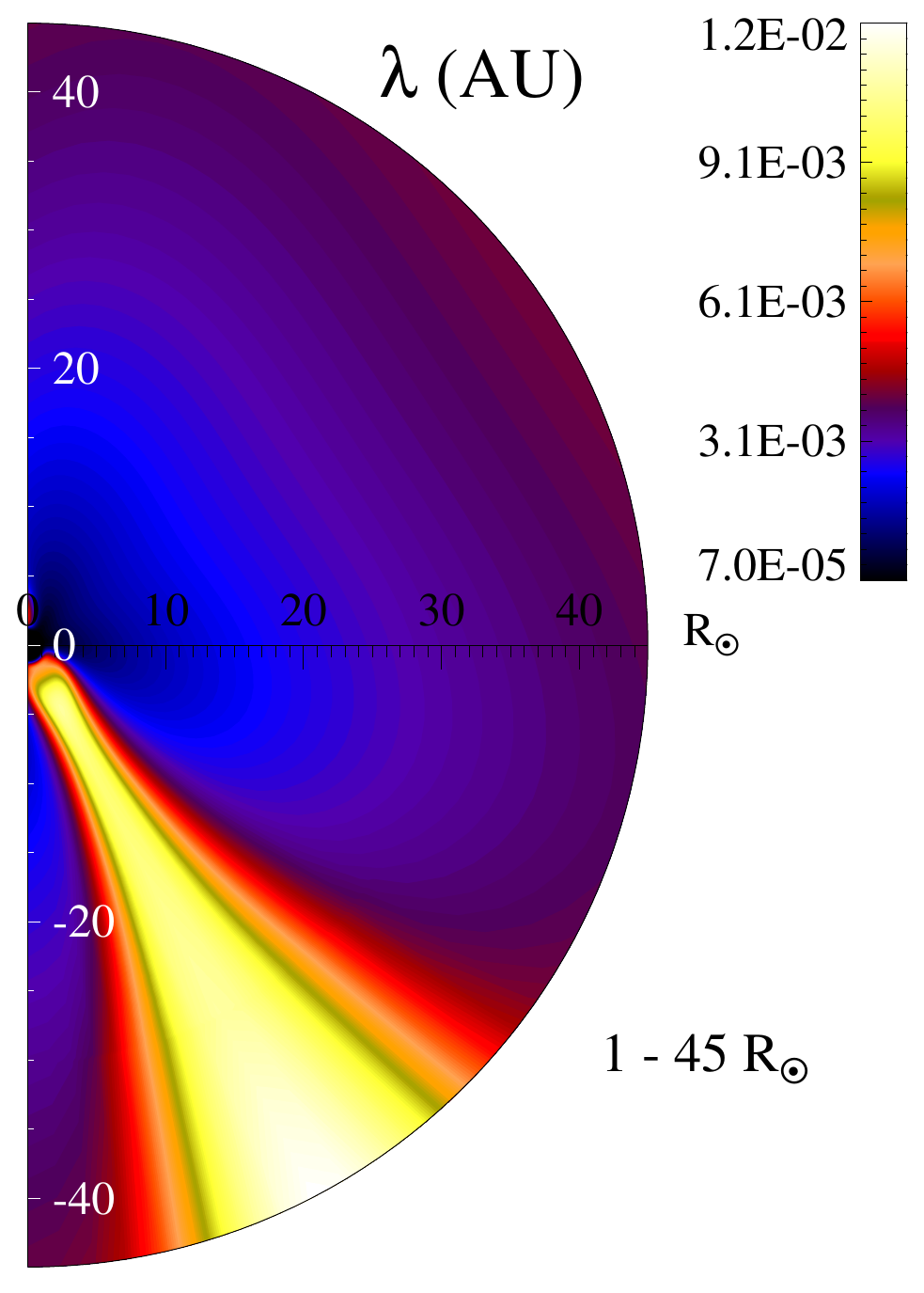}{0.25\textwidth}{}
          }
\gridline{\fig{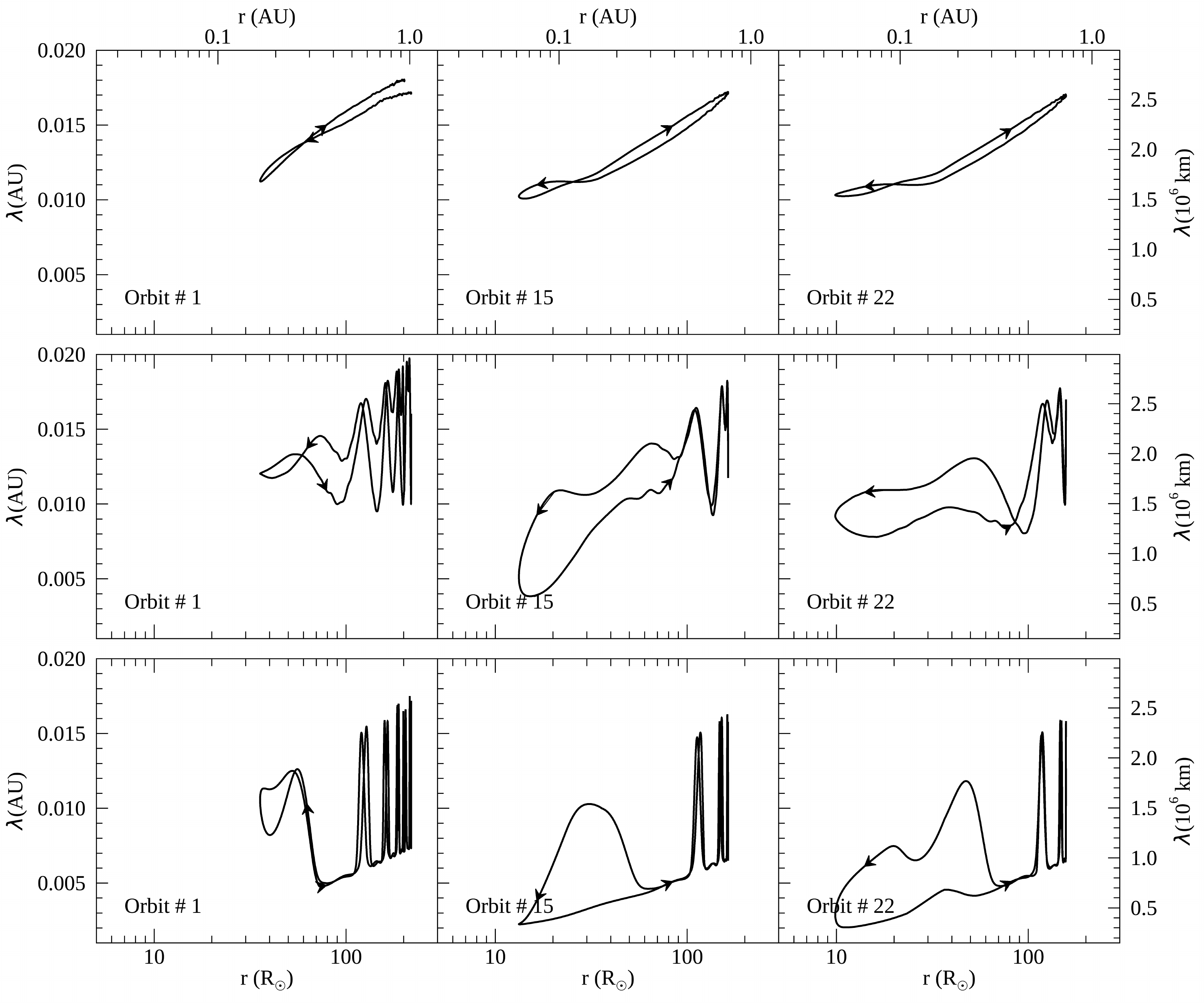}{.9\textwidth}{}}
\caption{Top panel shows correlation scale of fluctuations \(\lambda\) in a meridional plane in the region \(1\dash 45~\rs\), from simulations with source dipoles tilted by (left) 0\degree, (middle) 10\degree, and (right) 60\degree~relative to the solar rotation axis. The next three panels show \(\lambda\) along the \psp~trajectory for selected orbits, for the three dipole tilts. Direction of arrows indicates inbound/outbound sections of orbits.}
\label{fig:corr0}
\end{figure}

\begin{figure}
\gridline{\fig{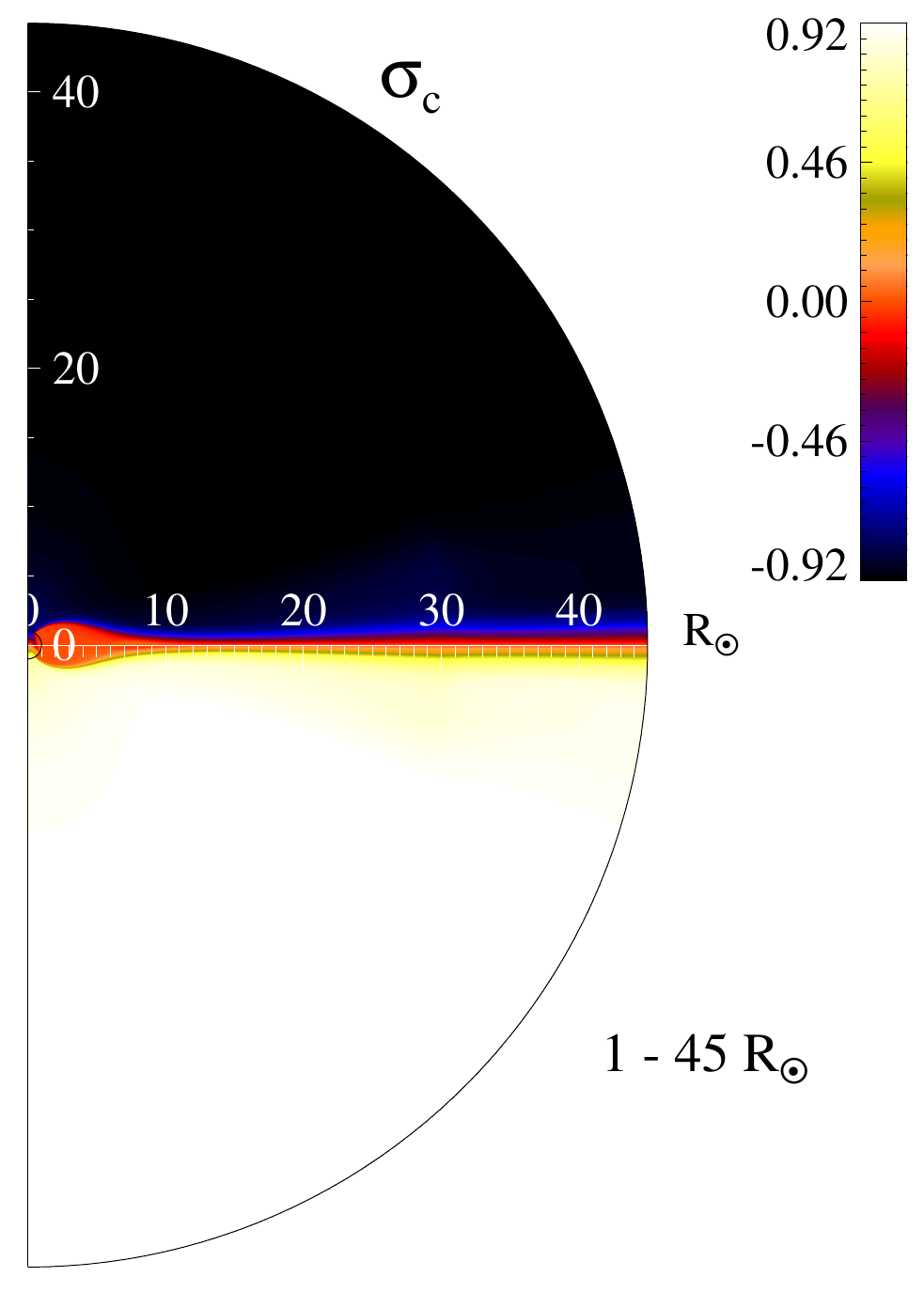}{0.25\textwidth}{}
          \fig{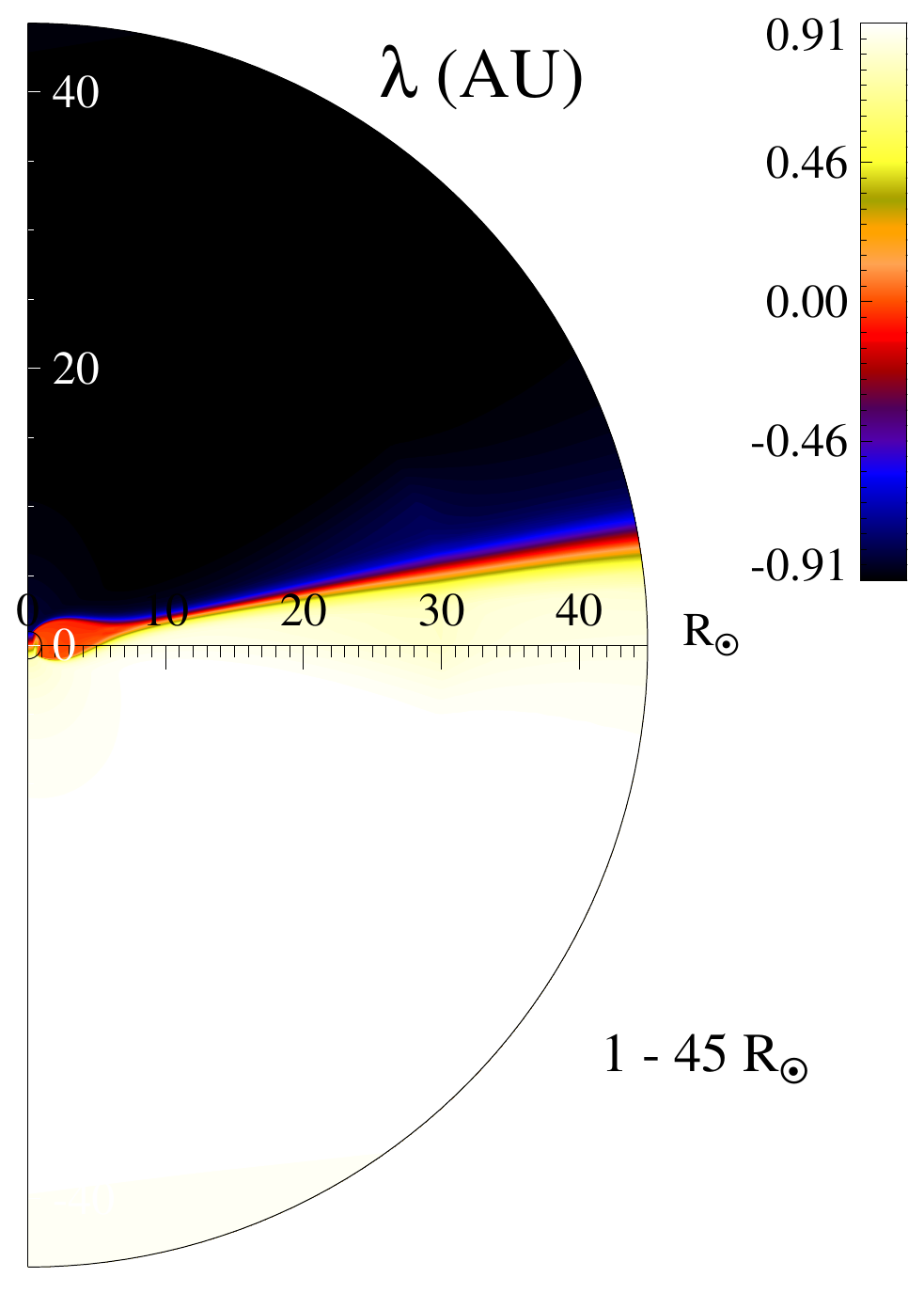}{0.25\textwidth}{}
          \fig{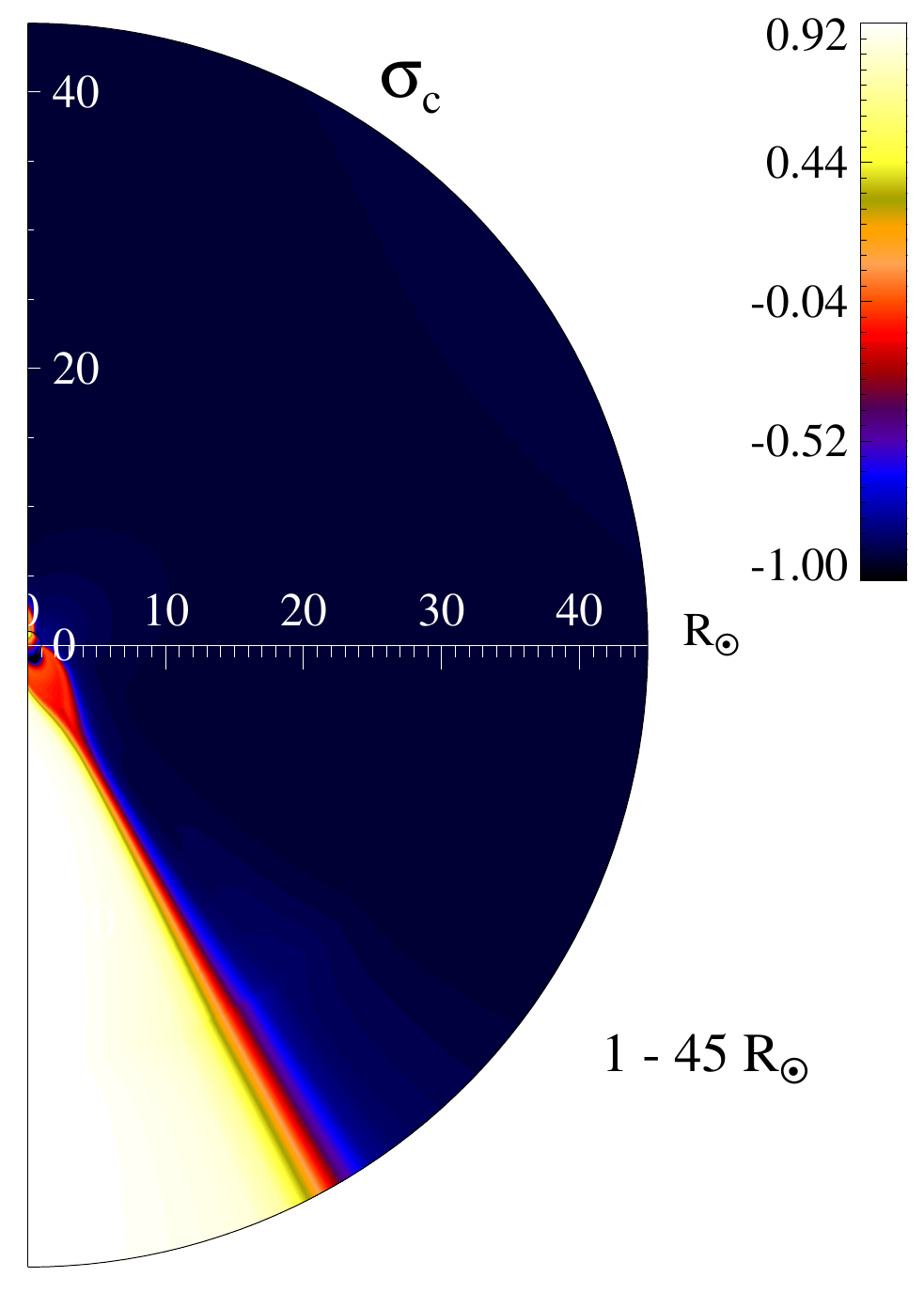}{0.25\textwidth}{}
          }
\gridline{\fig{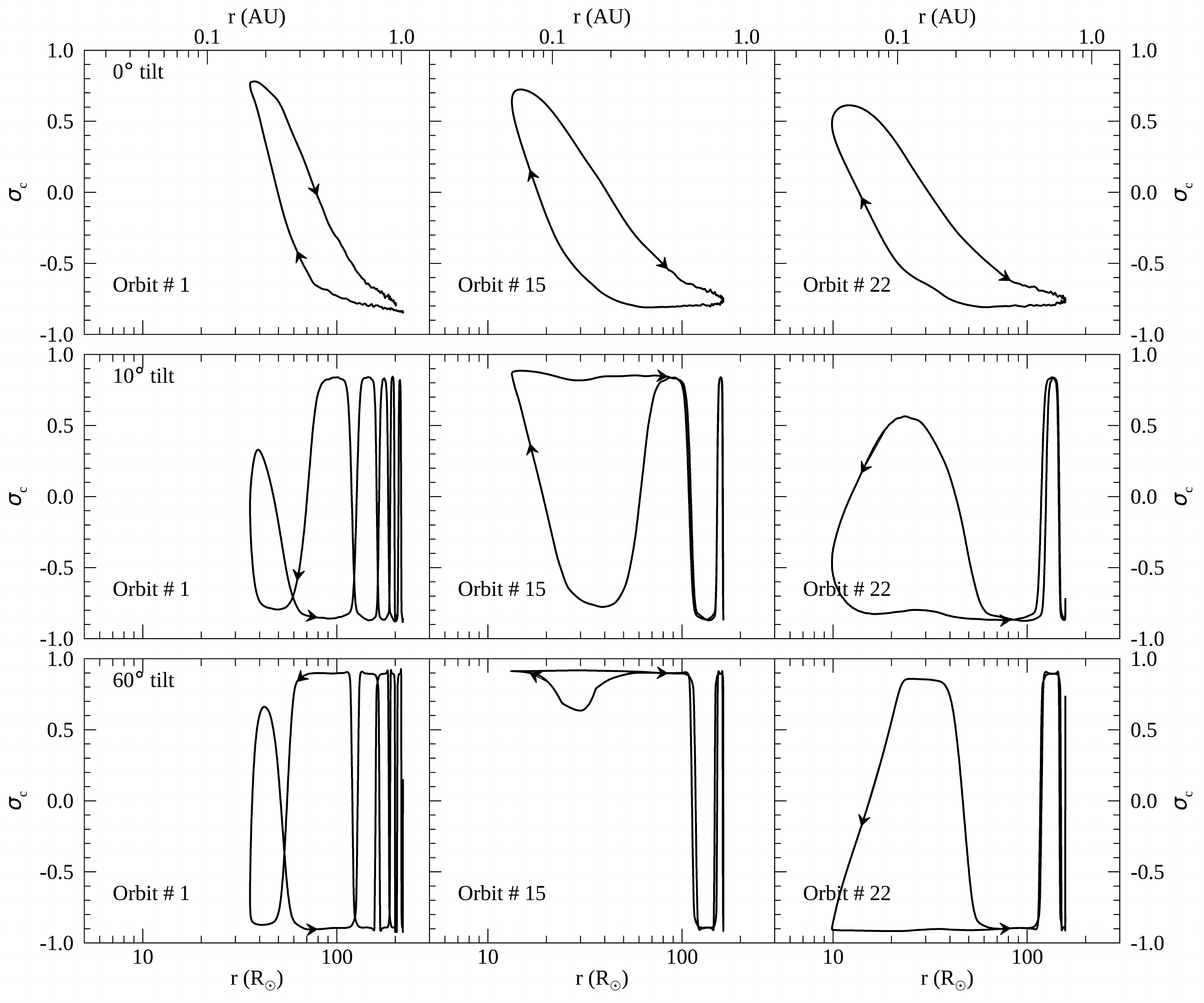}{.9\textwidth}{}}
\caption{Top panel shows normalized cross helicity \(\sigma_c\) in a meridional plane in the region \(1\dash 35~\rs\), from simulations with source dipoles tilted by (left) 0\degree, (middle) 10\degree, and (right) 60\degree~relative to the solar rotation axis. The next three panels show \(\sigma_c\) along the \psp~trajectory for selected orbits, for the three dipole tilts.  Direction of arrows indicates inbound/outbound sections of orbits.}
\label{fig:sigc0}
\end{figure}

As a first set of results from our three fiducial runs,
we extract data from the computed 
steady two-scale MHD solutions, and 
examine  the distribution of 
turbulence and plasma properties -- the 
(fluid velocity plus magnetic) fluctuation energy per unit mass, 
the cross helicity,
a single computed correlation scale,
the fractional magnetic fluctuation (i.e., ``delta B/B''),
and the plasma \(\beta\).  
The two top panels of Figure \ref{fig:z0} 
portray the distribution of turbulence energy density
in an arbitrarily chosen meridional plane, for 
tilt angles 0\degree, 10\degree, and 60\degree. 
The top panels of Figures \ref{fig:corr0} and \ref{fig:sigc0} show the corresponding distributions of the correlation scale \(\lambda\) and normalized cross helicity 
\(\sigma_c\). Note that the lower panels of these figures depict samples computed along trajectories, which will be described in the 
following section.

The three meridional plane panels in Figure \ref{fig:z0} 
show that the conditions near the ecliptic plane change considerably with 
increasing dipole tilt. For the unltilted case the 
regions of highest turbulence level are found exclusively at higher latitudes, 
and one can penetrate deeply into the 
corona near the ecliptic plane without encountering these
regions. For 10\degree~tilt the region of higher turbulence levels bulges out slightly at high latitudes 
and grazes the ecliptic plane region. At the highest tilt (60\degree) the ecliptic plane is fully engulfed in the region of higher fluctuation levels. 

Turning to the behavior of the correlation scale \(\lambda\), we focus 
attention on the top three panels of Figure \ref{fig:corr0}.
Here we can see that there is a general tendency for
the correlation length to grow with increasing heliocentric distance, as is well known 
from both observations in the inner heliosphere \citep{smith2001JGR,breech2008turbulence}
and turbulence theory \citep{karman1938prsl,hossain1995PhFl,Zank2017ApJ835}.
It is also clear that the behavior of \(\lambda\) is very different 
at low latitudes at solar minimum (0\degree~tilt),
and more generally in the vicinity of the heliospheric current sheet (HCS) for all tilt angles.  

Finally, the top three panels of 
Figure \ref{fig:sigc0}
illustrate the behavior of the normalized cross helicity 
\(\sigma_c\) in meridional planes. 
At zero tilt, almost all latitudes 
remain at high cross helicity, the sign being associated 
with outward propagation, and therefore reversing 
across the HCS, 
where the large scale magnetic polarity changes sign. 
However, the narrow region near the low-latitude HCS behaves differently. Within about 5 to 8 \(\rs\)
a region of low cross helicity is formed in steady state, 
associated with low-latitude closed field lines 
(coronal streamers) that experience Alfv\'enic 
propagation from the inner boundary in 
\textit{both} directions.
This region narrows at the top of the streamers, and then 
very gradually widens towards
increasing heliocentric distances. 
Note that the regions depicted extend only to 45 $\rs$ and 
therefore not yet seen is 
the general tendency for decrease of cross helicity due to expansion
\citep{zhou1989grl,usmanov2014three}, and the more rapid, localized decrease due to shear driving \citep{roberts1992jgr,breech2008turbulence}. The latter effect may possibly be not fully-accounted for in the present simulations \citep{usmanov2018}, which lack microstream driving of turbulence \cite[see][]{breech2008turbulence}.
On the other hand, the only 
regions of near-zero cross helicity seen in these simple tilted-dipole simulations is the region within a few degrees of the HCS.
It will be interesting to see if \psp~passes though more widely 
distributed lower cross helicity regions 
in orbits during solar maximum when the 
HCS might be more disordered than a tilted dipole. 
\subsection{Turbulence Parameters along \psp~Trajectory}
An entirely different view of the 
state of the heliospheric plasma is afforded
by sampling along the trajectory of the \psp~spacecraft.
Here we employ the 
same three datasets as above, 
at varying dipole tilt, 
but in this case 
sampled along the anticipated \psp~trajectory (extracted from a \href{https://naif.jpl.nasa.gov/naif/index.html}{NASA  \textit{SPICE} kernel}) for selected orbits, taking solar rotation into account.

This provides a plausible scenario for the pattern of variations that the 
mission will experience in different orbits at 
different phases of the solar cycle.
These results are shown in the lower panels 
of Figures \ref{fig:z0} -- \ref{fig:sigc0} 
for the 
same 
three turbulence quantities -- turbulence energy density \(Z^2\), correlation scale \(\lambda\), and cross helicity \(\sigma_c\). 
Sampling these
turbulence properties 
for three levels of dipole tilt 
enables an estimation of 
variation due to anticipated rising level of solar activity. 

The figures suggest that the \psp~will encounter an increased \(Z^2\) as it approaches the region where turbulent fluctuations are generated \citep[e.g.,][]{matthaeus1999ApJL523}. The turbulence is less ``aged'' in these regions \citep{matthaeus1998JGR}, however, and therefore the correlation scale is expected to decrease as the spacecraft approaches its perihelia. Note that the trajectory plots have two ``lobes'', since the inbound and outbound trajectories are not identical. The lobes intersect as the HCS is crossed.

The turbulence level (Figure \ref{fig:z0}) seen on orbit 1 is generally 
lower than orbits 15 and 22, for the untilted dipole case, 
mainly because the perihelion is 
lower in the latter two cases.
As we move towards higher dipole tilts, still higher 
turbulence levels are seen, punctuated by 
relative sudden drops in the level, due to \psp~orbital 
crossings of the current sheet region. 
The very high levels of turbulence experienced in orbit 22, in a 60\degree~tilted dipole, 
are due to the spacecraft penetrating deeply into 
the lobes of higher turbulence 
levels found far from the HCS. In the same orbit there 
remain a few HCS crossings, characterized
by brief periods of lower \(Z^2\).

Note that the specific pattern of HCS crossings is determined by the intial (launch) heliolongitude of the \psp, which is arbitrarily placed within the simulation for the purposes of the present study. It is possible to vary this initial longitude and  perform an average over the different trajectories so obtained, as was done in \cite{chhiber2019psp1} to estimate the time spent by \psp~within various critical surfaces. However, this procedure would smooth out the large variations associated with HCS crossings, which we find worth preserving in our presentation here. However, we do employ such an ensemble of trajectories to investigate trends in the cross helicity measured by the simulated spacecraft (see below).

The correlation length estimates shown in 
Figure \ref{fig:corr0} also show systematic variations
along \psp~orbits at different stages of solar activity.
The general increase of \(\lambda\) with increasing 
heliocentric distance is most evident in the 
orbital sampling for the untilted dipole case. 
Here one also sees a slight flattening of the 
 variation of \(\lambda\) for orbits 15 and 22
 that descend significantly below 25 \(\rs\).
For greater solar activity and greater dipole tilts, the 
behavior of correlation length along the orbits is much more erratic,
punctuated by large excursions near HCS crossings.
There are also significant excursions inside of 
25 $\rs$ associated with passage more deeply 
into lobes with different levels of turbulence activity.
Under typical circumstances the correlation scales grows
with increasing turbulence age \citep{matthaeus1998JGR}, so the excursions of $\lambda$ seen along orbits at higher solar activity may be thought of 
as alternately sampling ``older'' and ``younger'' turbulence.
These variations of correlation scale may have immediate implications for variation of energetic particle diffusion coefficients, which nominally scale in proportion to an outer scale of
the fluctuations \citep{jokipii1966cosmic,chhiber2017ApJS230,zhao2017ApJcr1}.
Note that the turbulence amplitude \(Z^2\) is smaller in the HCS, in essentially the same locations as those in which correlation scale \(\lambda\) is larger -- as suggested above, this is  indicative of ``older turbulence''.

The cross helicity \(\sigma_c\) also varies in interesting ways 
along the \psp~orbits, as shown in Figure \ref{fig:sigc0}.
An asymmetry during inbound and outbound orbital segments
is seen in the zero tilt case, which translates into a
greater part of the inbound orbit spent in very highly Alfv\'enic plasma, 
as compared to the outbound leg of the same orbit. 
For larger tilt angles one also finds several periods of time in 
which the spacecraft is located in highly Alfv\'enic solar wind,
an effect that 
can occur during inward or outward segments.
Another notable feature is again the 
rapid changes associated with 
HCS crossings. 
Like the turbulence energy and the correlation scale, 
these rapid changes of cross helicity occur mainly 
beyond 100 \(\rs\).

\begin{figure}
\gridline{\fig{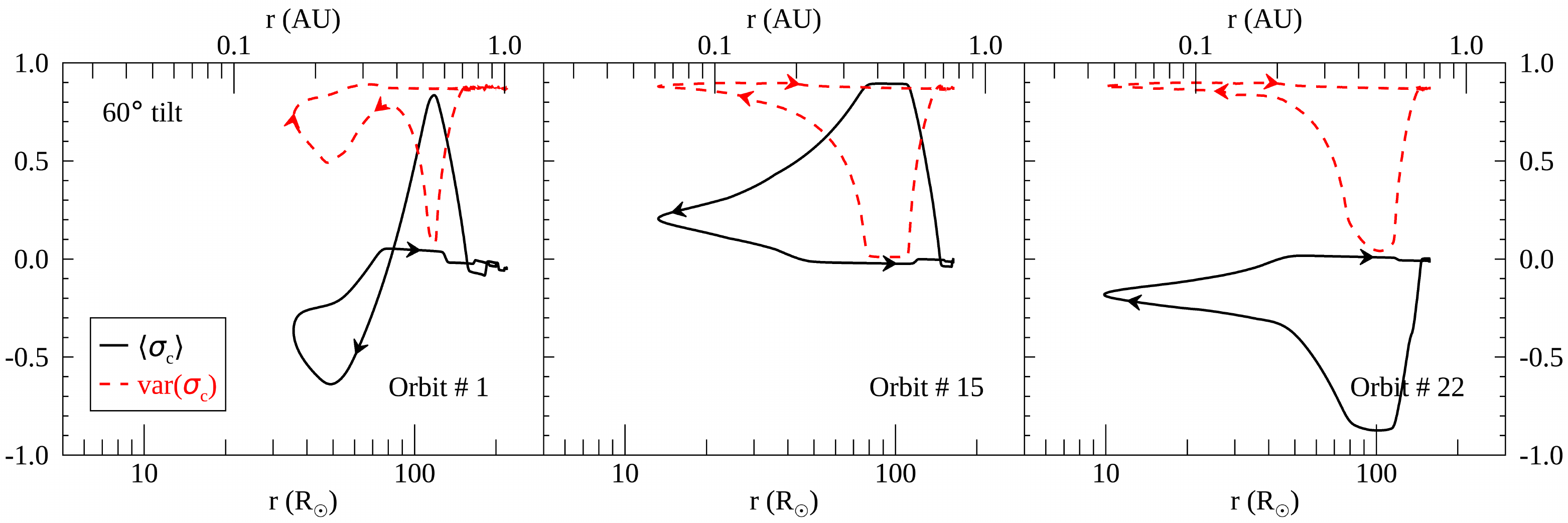}{.95\textwidth}{}}
\caption{Mean cross helicity \(\langle\sigma_c\rangle\) (solid black curve) and standard deviation of cross helicity var(\(\sigma_c\)) (dashed red curve) of cross helicity for selected \psp~orbits in the 60\degree~dipole-tilt run, computed from an ensemble of trajectories obtained by varying the initial (launch) heliolongitude (see text).}
\label{fig:sigc0_stat}
\end{figure}    
To further investigate the cross helicity measured along the simulated trajectory in the 60\degree~case, we vary the initial heliolongitude of the trajectory and perform a statistical analysis. We consider \(\sim 100\) values of the initial longitude \(\phi_{\psp,0}\), ranging from  0\degree~to 359\degree, and perform an average over them. That is, we first find \(\sigma_c\) along \textit{each} \psp~trajectory defined by a value of \(\phi_{\psp,0}\). We then average over the different \(\phi_{\psp,0}\) to obtain a \textit{mean} \(\sigma_c\), plotted using a solid black curve in Figure \ref{fig:sigc0_stat}. The dashed red curve shows the standard deviation of \(\sigma_c\) computed over the different trajectories. In the statistical ensemble of \(\sim 100\) trajectories obtained by varying the launch longitude, these results suggest that \psp~is likely to see large fluctuations in cross helicity in outbound section of its orbits, relative to the inbound section. This difference presumably arises in the geometrical asymmetry between the inbound and outbound sections of the trajectory.

\begin{figure}
\centering
\includegraphics[scale=.5]{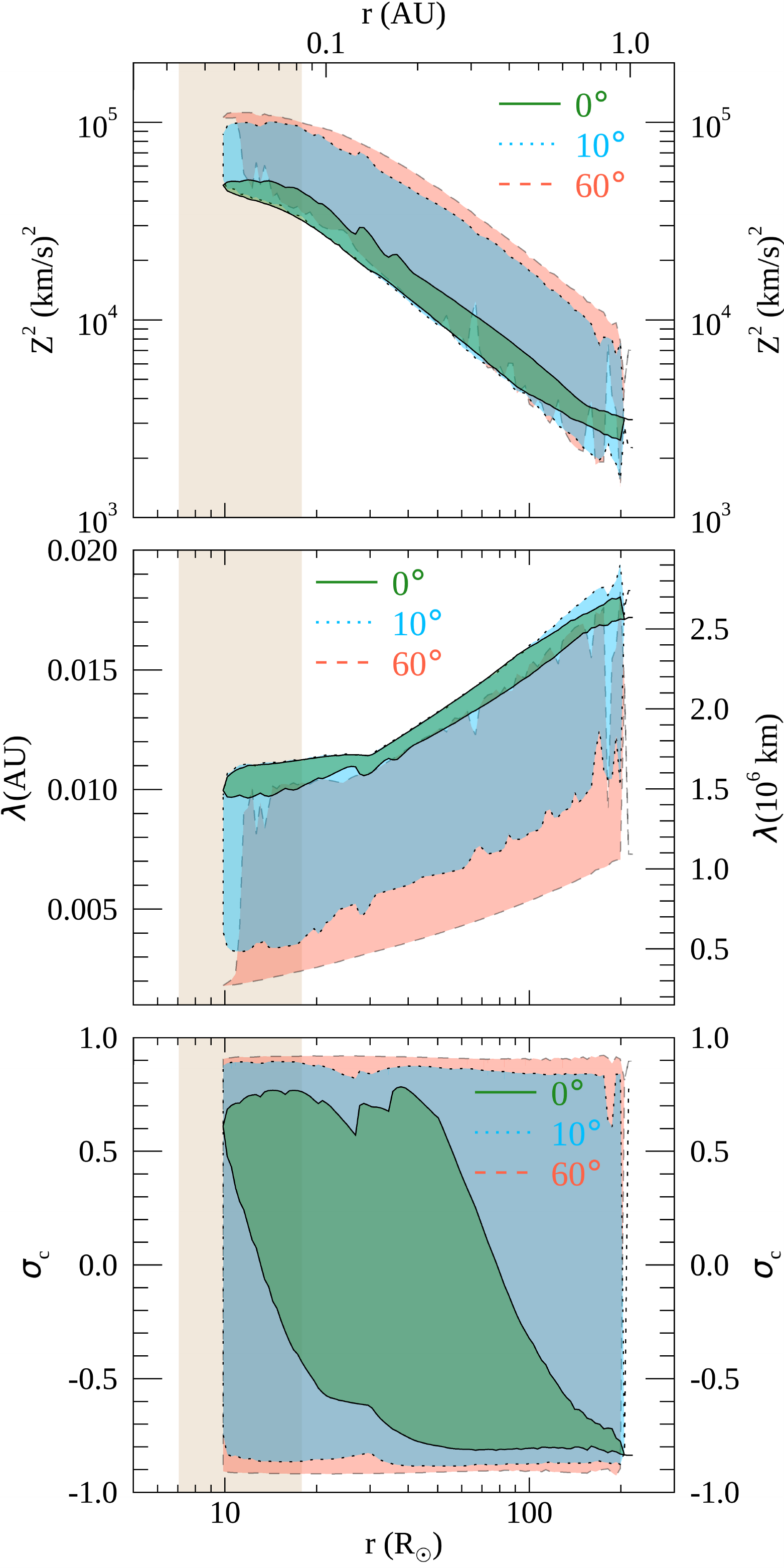}
\caption{Bands showing the range of values of turbulence parameters (top: turbulence energy; middle: correlation scale; bottom: cross helicity) encountered during all orbits, for simulations with source dipoles tilted by 0\degree (green; solid border), 10\degree (blue; dotted border), and 60\degree~(red; dashed border) relative to the solar rotation axis. The tan shaded vertical band demarcates the location of the Alfv\'en surface.}
\label{fig:bands}
\end{figure}  

A different view of the turbulence properties 
that \psp~is likely to encounter is provided in 
Figure \ref{fig:bands}.
Here the content of Figures \ref{fig:z0}--\ref{fig:sigc0}
is summarized by plotting the 
\textit{range} of values of \(Z^2, \lambda\), and \(\sigma_c\)
spanned during all orbits; the color coding 
indicates the different runs with varying dipole tilt, 
and the shaded area on the left shows 
the range of values of heliocentric radius at which the 
Alfv\'en surface is found at different heliolatitudes, thus defining a type 
of Alfv\'en critical region \citep{chhiber2019psp1}. 
This compilation of data illustrates
clearly that \psp~will encounter the narrowest 
range of turbulence parameter values when 
orbiting through a solar minimum state with 0\degree~tilt. 
Conversely the solar maximum proxy, a state with 60\degree~tilt, 
sets the stage for encountering the widest range of turbulence 
conditions. Overall, one might conclude, in rough terms, since
the \psp~orbits will likely encounter both 
minimum and maximum activity periods, 
that the turbulence energy may vary by a factor of 5 at any given 
heliocentric radial distance. Meanwhile the 
correlation scale may vary by a factor of three or so. 
Regions with widely varying normalized cross helicity 
will be encountered throughout the mission, 
although very orderly solar minimum conditions
give rise to orderly observations of cross helicity, 
as expected. 
\begin{figure}
\centering
\includegraphics[scale=.55]{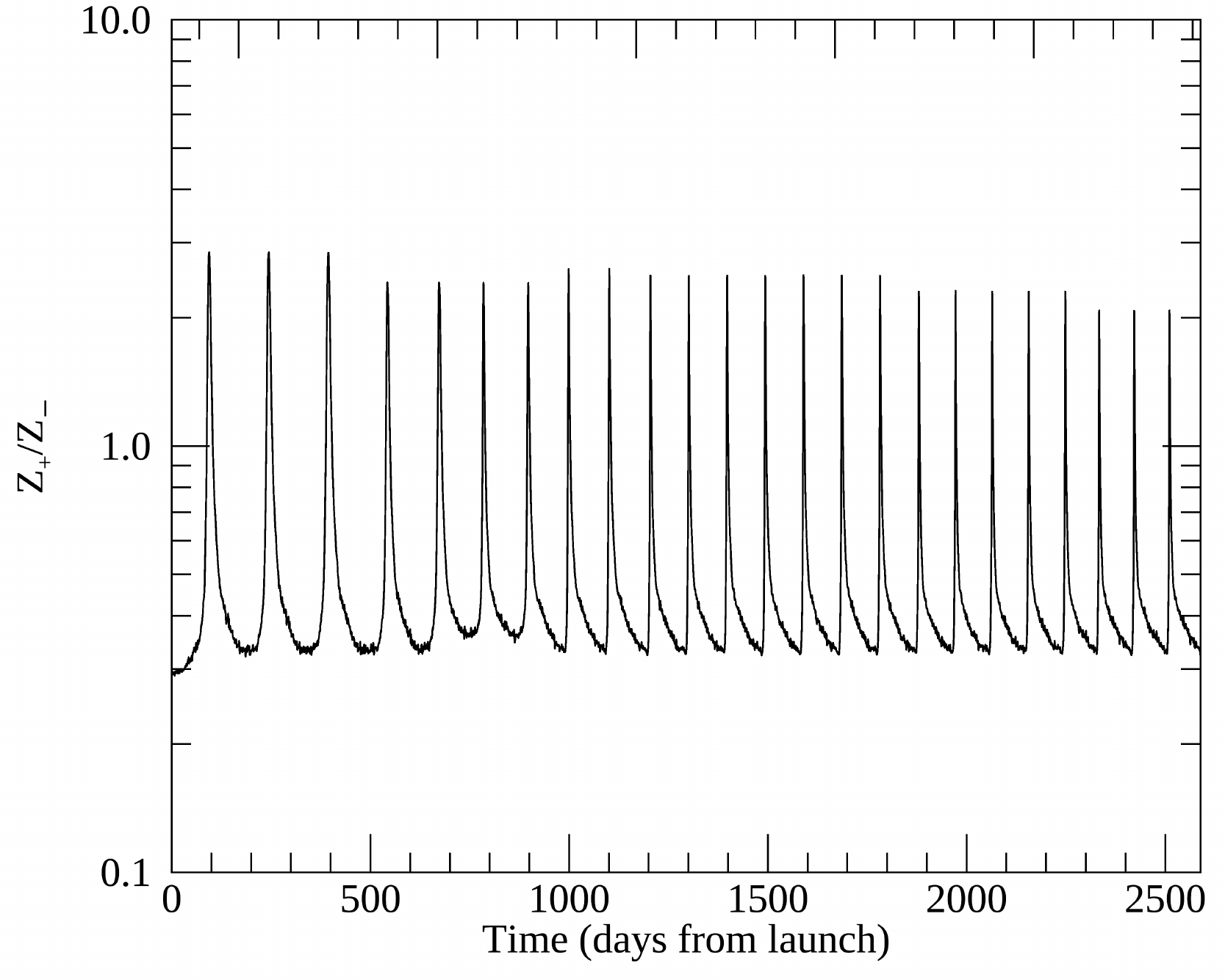}
\caption{Relative strength of the two Els\"asser modes along the \psp~trajectory, from an untilted dipole simulation.}
\label{fig:zpm0}
\end{figure}    

Closely related to the cross helicity is the 
examination the relative strength of ``inward'' and ``outward'' modes, defined in terms of the Elsasser variables \(\bm{z}_\pm = \vb' \pm \bm{b}'\) \citep{elsasser1950PhRv}. Using the identity \(Z_\pm^2 = (1 \pm \sigma_c) Z^2\), where \(Z_\pm^2 = \langle\lvert \bm{z}_\pm\rvert^2 \rangle\), we plot the ratio \(Z_+/Z_-\) throughout the \psp~trajectory in Figure \ref{fig:zpm0}, for the untilted dipole case. Once again, we see that the orbits will cross from regions of dominant \(\bm{z}_-\) to those where \(\bm{z}_+\) is dominant. Note that in the simulation considered here, the ``outward'' propagating mode is \(\bm{z}_-\) in the Northern solar hemisphere (where the magnetic field points radially), while \(\bm{z}_+\) propagates outward in the Southern hemisphere.
\begin{figure}
\centering
\includegraphics[scale=.55]{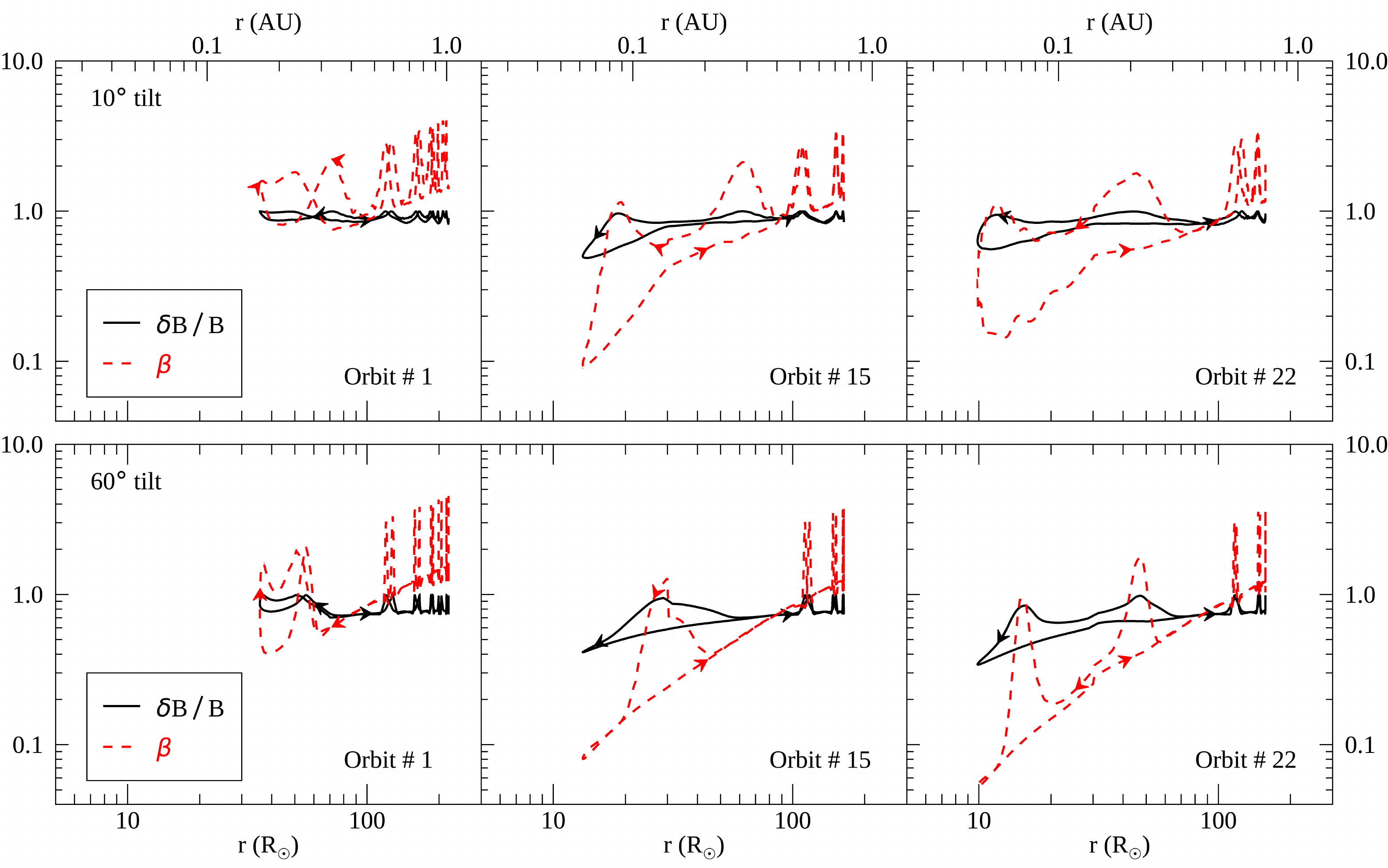}
\caption{Relative fluctuation strength and plasma \(\beta\) along \psp~trajectory for dipoles with 10\degree~(top) and 60\degree~(bottom) tilt relative to the solar rotation axis.}
\label{fig:beta}
\end{figure}    

Figure \ref{fig:beta}
illustrates, for the 10\degree~and 60\degree~dipole tilt runs, 
and for several of the \psp~orbits, 
the variation of two important quantities for plasma physics considerations, namely 
\(\delta B/\mathcal{B}\) and plasma \(\beta\). 
In the usual way \(\delta B\) is defined as the root mean square (rms) magnetic fluctuation amplitude (here \(\langle B'^2 \rangle^{1/2}\)), and \(\mathcal{B}\) denotes the average (rms) local field strength, derived from both the resolved large-scale field and the mean value of the turbulence energy: \(\mathcal{B} = (B^2 + \delta B^2)^{1/2}\). To estimate \(\delta B\), we first convert $Z^2$ to $\langle B'^2\rangle$ using the definitions \(Z^2 = \langle v'^2 + b'^2\rangle\) and \(\bm{b}' = \bb'(4\pi\rho)^{-1/2}\): \(\langle B'^2 \rangle = 4 \pi \rho Z^2/(r_\text{A}+1)\), where $r_\text{A} = \langle v'^2 \rangle /\langle b'^2 \rangle$ is the Alfv\'{e}n ratio, here taken to be equal to 1/2 for consistency with the constant \(\sigma_D=-1/3\) used in our model (Section \ref{sec:model}).\footnote{\footnotesize{The value of the Alfv\'en ratio is expected to increase from \(\sim 1/2\) near Earth to \(\sim 1\) in the near-Sun environment, according observations by \textit{Ulysses}. We have checked that the results presented here do not change significantly if we set \(r_\text{A}=1\), and we therefore use \(r_\text{A}=1/2\) for consistency with the constant \(\sigma_D=-1/3\) used in our model. Note that some recent models of turbulence transport in the solar wind feature a dynamical equation for the energy difference \citep{Zank2017ApJ835,zank2018ApJ}, and it would be interesting to compare the present results with predictions based on such models.}} Plasma \(\beta\) is the ratio of gas pressure to magnetic pressure: \(\beta = (P_S + P_E)/P_M\). Here \(P_S\) and \(P_E\) are the proton and electron pressures respectively, and the magnetic pressure is \(P_M = \mathcal{B}^2/(8\pi)\). We can see on this illustration, particularly for the plasma 
\(\beta\), that larger variations are seen in later orbits that probe lower altitudes. During solar-max like conditions (bottom panel), the plasma \(\beta\) reaches values as low as \(\sim 0.06\) during the perihelia of later orbits. Low values of \(\beta\) provide justification for a highly anisotropic nearly two-dimensional (2D) representation of turbulence in the inner corona \citep{matthaeus1990JGR,zank1993nearly,zank2018ApJ}. Note that the ``spikes'' in the value of \(\beta\) arise when \psp~crosses the HCS. Greater variation in \(\delta B/B\) is also seen in later orbits. For the 60\degree~dipole case, this ratio is about 0.3 during the final perihelia.
\subsection{Validity of Taylor Hypothesis along \psp~Trajectory}  
Spacecraft observations generally take the form of single-point (in space) time series of data. Time-lagged correlation data based on this single-spacecraft signal
can be interpreted as spatially-lagged correlation data 
if the sampled structures in the observed signal are 
swept past the detector rapidly enough that 
they experience negligible distortion during their transit. 
Achievement of such a condition 
requires the speed of convection past the spacecraft (determined by the velocities of the wind and the spacecraft) to be much larger than the characteristic speed of dynamical 
interactions, especially nonlinear interactions. 
The standard Taylor ``frozen-in'' approximation \citep{taylor1938ProcRSL}, 
also known as the Taylor Hypothesis (TH),   
is useful \citep[e.g.,][]{matthaeus1982JGR,chhiber2018MMS} in the 
supersonic and super-Alfv\'enic solar wind 
that is typically encountered 
by spacecraft near Earth, \textit{if} the dynamical process 
of interest can described at the MHD level.\footnote{\footnotesize{Kinetic-scale activity 
may have timescales shorter than the convection timescale; in 
that case the validity of the frozen-in approximation may be questioned even near Earth \citep[see][]{howes2014ApJ,perri2017ApJS,chhiber2018MMS}.}}

At \psp~perihelia, especially in later orbits, the above conditions may not hold true. Moving towards lower heliocentric distances,
the Alfv\'en speed increases even as the wind speed decreases.
Wind and Alfv\'en speeds become equal at the 
Alfv\'en critical surface (or region), which  
is expected to be in the range of 10 -- 30 $\rs$ \citep[e.g.,][]{cranmer2007ApJS,
verdini2010ApJ,deforest2014ApJ787,perri2018JPP,chhiber2019psp1}.  
Therefore the standard Taylor hypothesis is not expected to 
apply for all regions to be explored by \psp. 

\begin{figure}
\centering
\includegraphics[scale=.55]{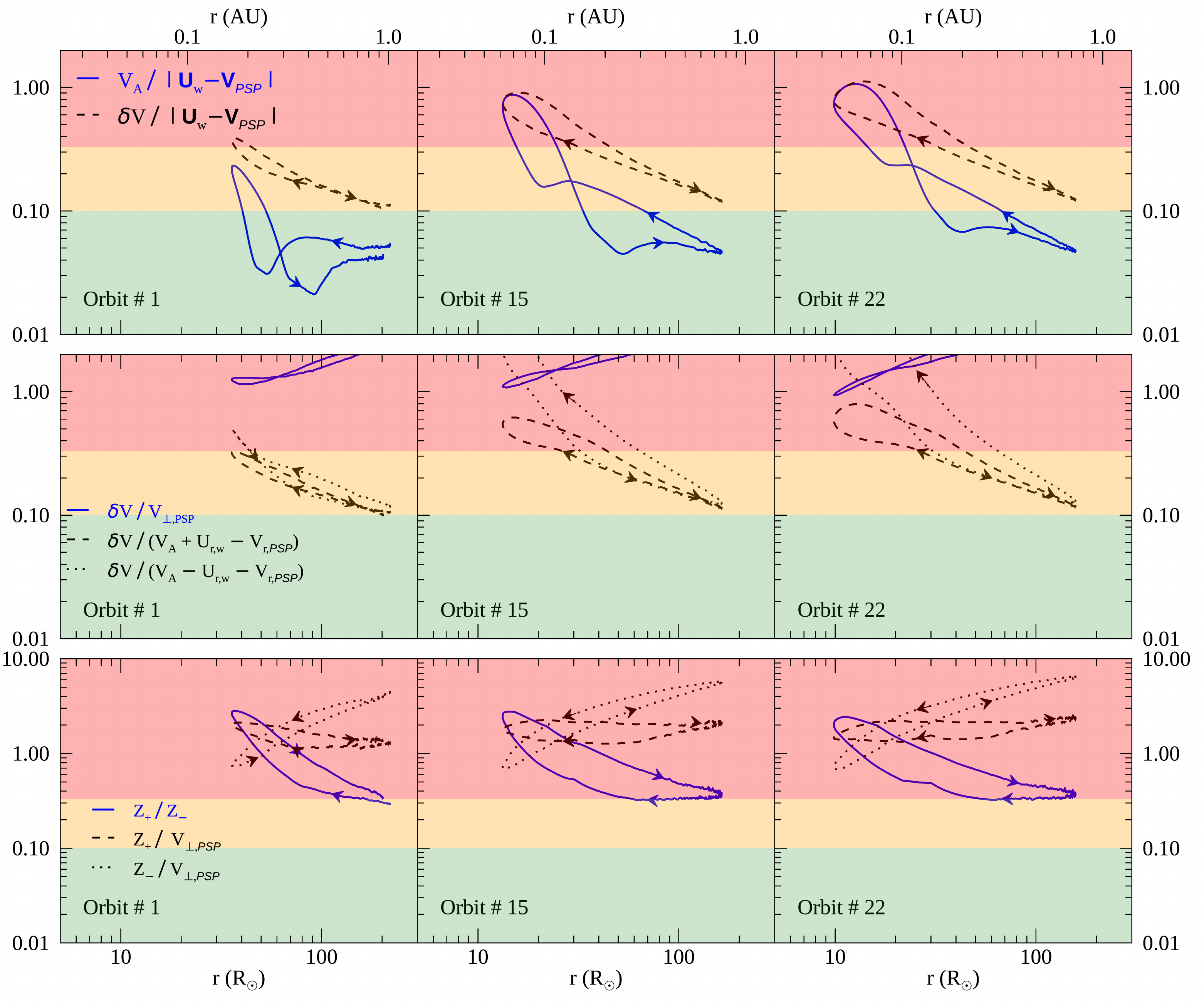}
\caption{Results from an untilted dipole simulation. Top: The plotted ratios compare the transit speed of the solar wind plasma in the \psp~frame \(\lvert \bm{U}_w - \bm{V}_\psp \rvert\) to the Alfv\'en speed $V_\text{A}$ (solid blue curve) and the characteristic speed of turbulent distortion $\delta V = Z$ (dashed black curve). Directions of arrows indicate ingoing and outgoing parts of the \psp~trajectory. Regions shaded green, orange, and red represent, respectively, high ($\text{ratio}<0.1$), moderate ($0.10<\text{ratio}<0.33$), and low ($\text{ratio}>0.33$) degrees of validity of the Taylor hypothesis. Middle and bottom panels: Tests of ``modified'' Taylor hypotheses along the \psp~trajectory (see text).}
\label{fig:taylor0}
\end{figure}

\begin{figure}
\centering
\includegraphics[scale=.55]{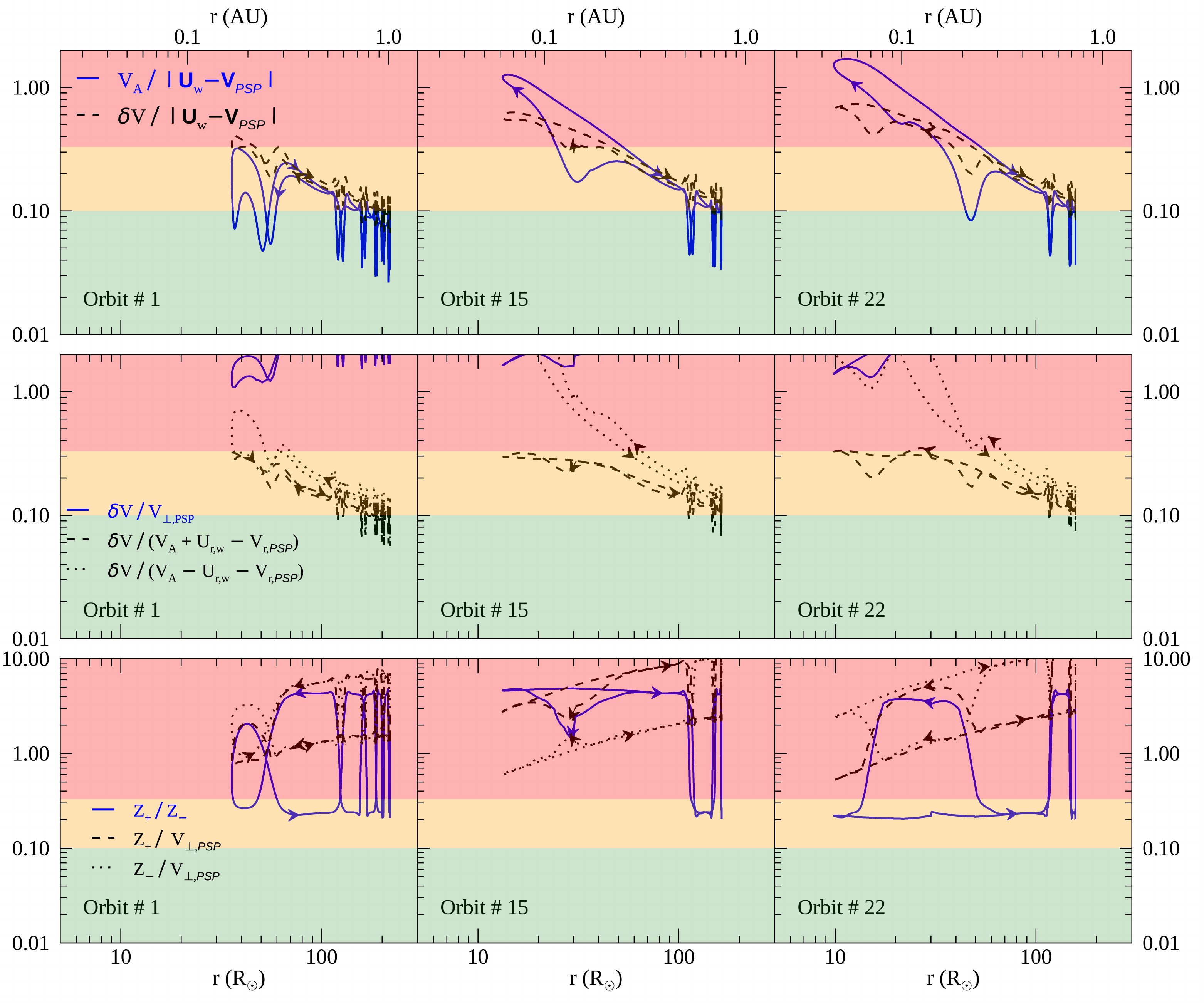}
\caption{Results from a simulation with a dipole tilted by 60\degree~relative to the solar rotation axis. Description of figures follows from Figure \ref{fig:taylor0}.}
\label{fig:taylor60}
\end{figure}

To test for possible periods of validity of the TH, as well as possible 
periods of its violation,
we use an untilted dipole simulation (Figure \ref{fig:taylor0}) and a simulation with 60\degree dipole tilt (Figure \ref{fig:taylor60}) to plot the ratios \(V_\text{A}/\lvert \bm{U}_w - \bm{V}_\psp \rvert\) and \(\delta V/\lvert \bm{U}_w - \bm{V}_\psp \rvert\) along selected \psp~orbits, shown in the top panels of the figures. 
The first of these ratios measures the speed of Alfv\'en waves (\(V_\text{A}\)) against the speed of convection of plasma past the spacecraft \(\lvert \bm{U}_w - \bm{V}_\psp \rvert\), where \(\bm{U}_w\) is the velocity of the wind and \(\bm{V}_\psp\) is the \psp~velocity (extracted from a \href{https://naif.jpl.nasa.gov/naif/index.html}{NASA  \textit{SPICE} kernel}). The second ratio measures a characteristic turbulent speed \(\delta V\) (taken to be \(\sqrt{\langle v'^2 \rangle} = \sqrt{2Z^2/3}\), assuming an Alfv\'en ratio \(r_\text{A} = 1/2\) again) against the convection speed.
For more discussion of the time scales relevant to the TH, see, e.g., \cite{matthaeus1997AIPCtrajectory}, \cite{klein2015ApJtaylor}, and \cite{bourouaine2018ApJ}.

We (arbitrarily) consider the TH to have high validity when the above ratios are smaller than 0.1 (green-shaded region in Figures \ref{fig:taylor0} and \ref{fig:taylor60}); when the ratios lie between 0.10 and 0.33 (orange-shaded region) we consider the TH to have intermediate-level validity; ratios greater than 0.33 imply poor validity (red-shaded region). 

As seen in the top panels of Figures \ref{fig:taylor0} and \ref{fig:taylor60}, the TH has good validity near Earth (\(\sim 215~\rs\)), and moderate validity up to around \(50~\rs\), but below this height the validity of the classical TH is questionable, with the perihelia of the later orbits laying deep within the poor-validity regime. The dips in the blue curve occur because of the \psp~crossing the HCS where the vanishing magnetic field lowers the Alfv\'en speed; these excursions are more numerous in the tilted dipole case. Note that the validity for the nonlinear speed \(\delta V\) is better during the inbound part of the orbit, when the wind velocity and the \psp~velocity are opposed.

Modified versions of the frozen-in hypothesis have been proposed \citep{matthaeus1997AIPCtrajectory,klein2015ApJtaylor} for use with the \psp~at or near
its perihelia. We now evaluate several of these.

The magnetic field is mostly radial close to the Sun, and it is possible that the \psp~may sweep \textit{across} the mean field with a speed \(V_{\perp,\psp}\) to sample 2D fluctuations quickly enough that their intrinsic frequencies can be neglected compared to the reciprocal of the transit time \citep{matthaeus1997AIPCtrajectory,klein2015ApJtaylor}. Note that 2D fluctuations have wavevectors perpendicular to the mean magnetic field \citep[e.g.,][]{oughton2015philtran}. This variation of the frozen-in approximation is tested in the middle panel (blue curve) of Figures \ref{fig:taylor0} and \ref{fig:taylor60}, indicating poor validity. Here \(V_{\perp,\psp} = 
(V_{\theta,\psp}^2 + V_{\phi,\psp}^2)^{1/2}\), where $V_{\theta,PSP}$ and $V_{\phi,PSP}$ are the polar and azimuthal speeds, respectively, of the \psp~in a heliocentric inertial frame.
%
%

A second variation of the TH is motivated by the anticipated high Alfv\'en speeds near the Sun. It is possible that slab fluctuations (with wavevectors parallel to the mean magnetic field \citep{oughton2015philtran}) are convected past the spacecraft by Alfv\'enic propagation before nonlinear effects can distort them  \citep{matthaeus1997AIPCtrajectory}. The speed of convection in the \psp~frame will be different for outgoing and ingoing modes: \(V_\text{A} + U_{r,w} - V_{r,\psp}\) for the former and \(V_\text{A} - U_{r,w} - V_{r,\psp}\) for the latter. Here $U_{r,w}$ and $V_{r,PSP}$ are the radial speeds of the solar wind and the \psp, respectively. Note that this variation of TH is not relevant for non-propagating 2D fluctuations. 
This Alf\'ven-speed corrected Taylor hypothesis was discussed
and implemented in analysis of \textit{Helios} data by
\cite{goldstein1986JGR}.
The black curves in the middle panels of Figures \ref{fig:taylor0} and \ref{fig:taylor60} test 
this modification of TH, finding that it works somewhat reasonably for inward-propagating slab modes (dashed black curve), especially during the inbound part of the orbit.\footnote{\footnotesize{It is worth noting that these variations of TH would have been more successful in the original Solar Probe mission, which had a planned perihelion below \(4~\rs\) \citep[see][]{matthaeus1997AIPCtrajectory}.}}

Finally, we consider the modified TH of \cite{klein2015ApJtaylor}. Noting that the Els\"asser mode \(\bm{z}_\pm\) is convected by the oppositely signed mode \(\bm{z}_\mp\), Klein et al. argue that the frozen-in approximation may be valid near the Sun if outward propagating modes dominate (\(Z_+ \gg Z_-\), assuming \(\bm{z}_+\) is the outward mode) \textit{and} if \(V_{\perp,\psp}\) is much larger than the speed of convection \(Z_-\). The bottom panel of Figure \ref{fig:taylor0} plots the ratios \(Z_+/Z_-\) (blue curve) and \(Z_\pm/V_{\perp,\psp}\) (black curves). The blue curve indicates the relative dominance of the outward mode at perihelion. In a tilted dipole case (Figure \ref{fig:taylor60}), the dominant mode at perihelion can change for different orbits. If the relative dominance of the outward mode is substantial, then the validity of the Klein et al. modification can be assessed by examining the black curve corresponding to the ratio of the minority mode speed to \(V_{\perp,\psp}\). Examining the bottom panels of Figures \ref{fig:taylor0} -- \ref{fig:taylor60}, we conclude that the validity of this modified TH remains questionable at perihelia. This result is broadly consistent with the findings of \cite{bourouaine2018ApJ}. Two factors combine to produce this result. One, if the \psp~perihelia passes through a low cross-helicity region, then the outward mode is not significantly dominant relative to the inward mode. Two, if the spacecraft passes through a region of large cross helicity, then the turbulence energy also increases (compare Figures \ref{fig:z0} and \ref{fig:sigc0}), which implies that the inequality \(V_{\perp,\psp} \gg Z_\pm\) doesn't hold.

\begin{figure}
\centering
\includegraphics[scale=.56]{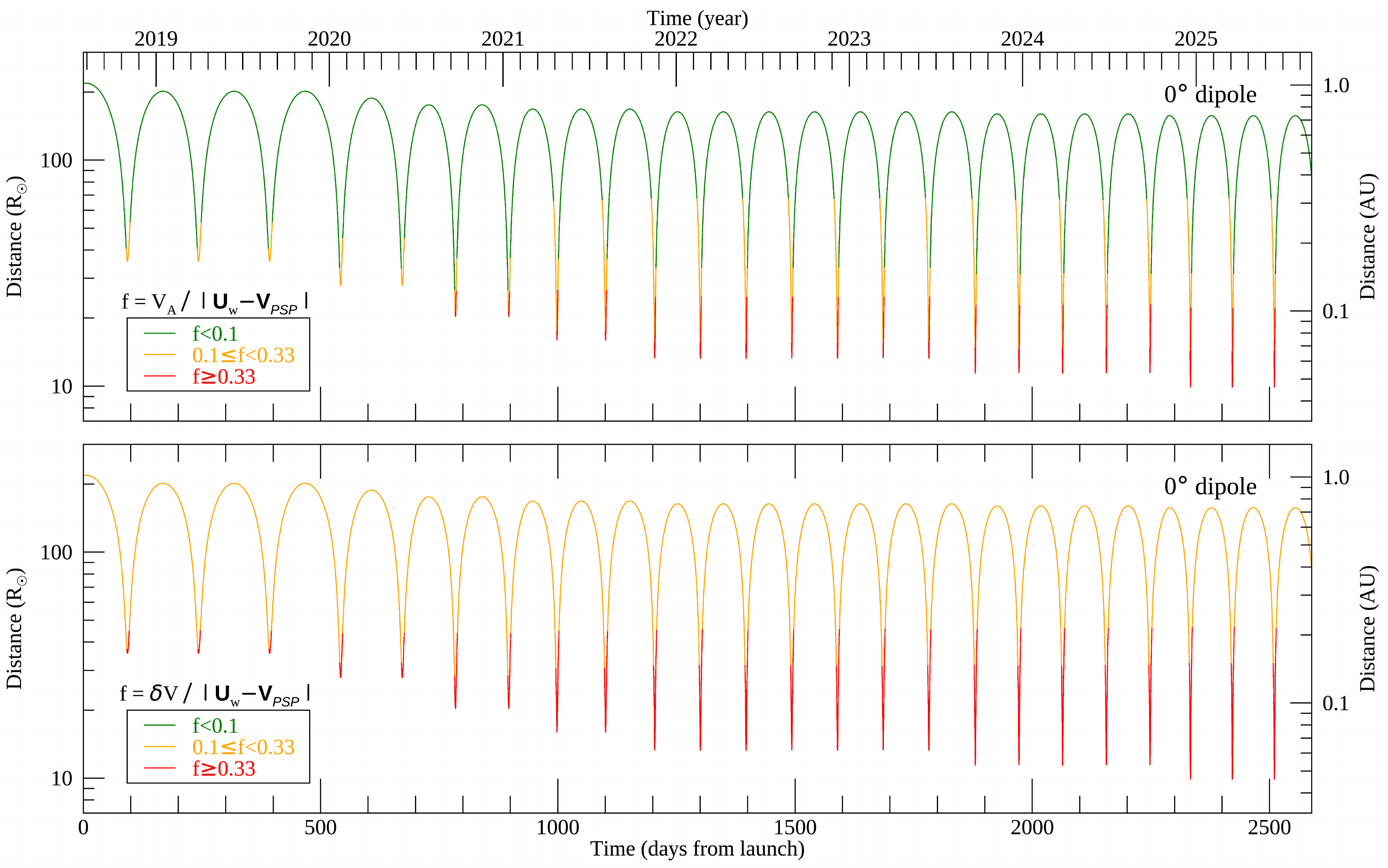}
\caption{Heliocentric radial position of the PSP, color-coded to indicate validity of the Taylor hypothesis during all orbits in the primary mission, computed from an untilted dipole simulation. The top and bottom panels examine the ratios \(V_\text{A}/\lvert \bm{U}_w - \bm{V}_\psp \rvert\) and \(\delta V/\lvert \bm{U}_w - \bm{V}_\psp \rvert\), respectively. Green, orange, and red segments of the orbits indicate high, intermediate, and low validity, respectively.}
\label{fig:taylor_full0}
\end{figure}

\begin{figure}
\centering
\includegraphics[scale=.56]{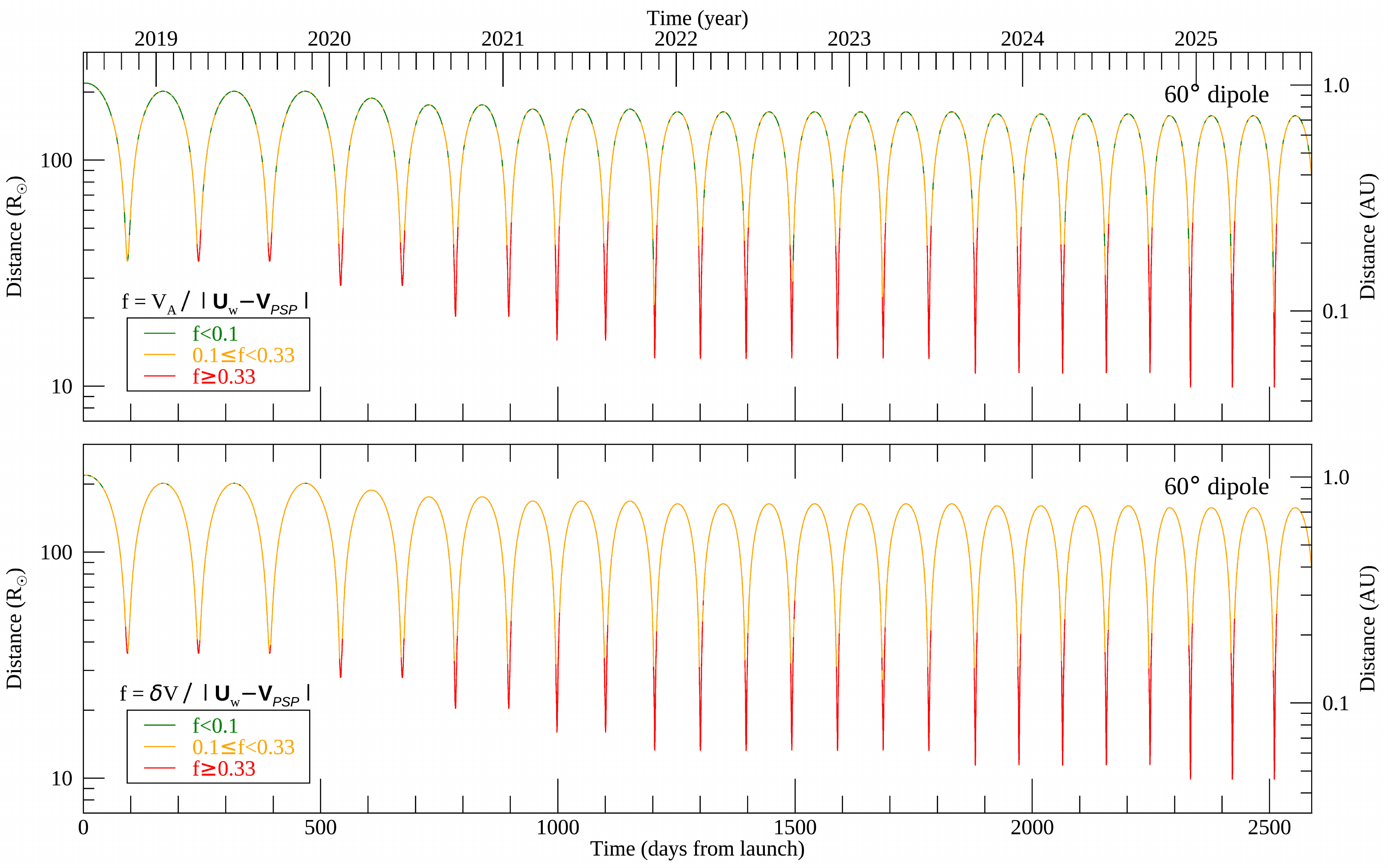}
\caption{Validity of the Taylor hypothesis over the enitire primary \psp~mission, computed from a simulation with source dipole tilted by 60\degree~relative to the solar rotation axis. The description of the figure follows from Figure \ref{fig:taylor_full0}.}
\label{fig:taylor_full60}
\end{figure}

Figure \ref{fig:taylor_full0} provides an illustration of the validity of the two standard forms of the Taylor hypothesis for all orbits in the the entire nominal \psp~mission, for the case of an untilted dipole. The two adopted forms of the TH are those respectively utilizing the turbulence speed or Alfv\'en speed, as discussed above,  compared with the combined transit speed due to the solar wind and the \psp~spacecraft speed. We can see here, for example, that the standard TH has no periods of high or even intermediate reliability inside of \(20~\rs\). Figure \ref{fig:taylor_full60} shows the same analysis, carried out for a 60\degree~tilted dipole simulation. Again, one finds no region of validity of the TH near any \psp~perihelion. This suggests that we will not be able to anticipate use of the TH and that different approaches to interpretation of \psp~data \citep[e.g.,][]{goldstein1986JGR,matthaeus1997AIPCtrajectory,
klein2015ApJtaylor,matthaeus2016prl,bourouaine2018ApJ} will need to be adopted in analyzing the most important periods of data acquisition inside of \(20~\rs\) and near perihelia. 

As an example of an alternative strategy of interpretation, there are expected to be interesting periods of near-corotation, known as fast radial scans \citep{fox2016SSR}. In such periods, the azimuthal speed of the \psp~in the Sun's rotating frame will be smaller than 1 km/s for a total of about 80 hours during the primary mission. Since the near-Sun plasma is expected to be in a state of near-corotation with the Sun \citep{weber1967ApJ148}, these will be times when the spacecraft could potentially take measurements in the frame of a parcel of non-propagating 2D turbulence. Such ``corotation'' intervals may then provide opportunities to study the time-evolution of the 2D turbulence. 
\section{Conclusions and Discussion}
This study 
is based on 
the implementation of 
a state of the art 3D numerical model
of the inner heliosphere, consisting of
Reynolds-averaged 
compressible MHD equations with separate proton and electron internal energy equations, and a turbulence model that is solved self-consistently with 
mean-flow equations
\citep{usmanov2014three,usmanov2018}. 
The model is employed to assess possible 
profiles of turbulence properties that 
might be seen by the recently launched
Parker Solar Probe mission. 
The several simulations employed for these 
context predictions are driven by boundary conditions consisting of a tilted dipole magnetic field.
This approach enables the simulation of conditions at least roughly corresponding to a range of solar activity that the Probe is likely
to encounter. These assessments are not intended to be specific predictions, but rather as guidelines for anticipation of ranges of conditions and their variability. 

We focused on only three turbulence parameters -- the 
energy density (per unit mass), the correlation scale, 
and the cross helicity (or degree of Alfv\'enicity). Our results suggest that \psp~is likely to measure increased turbulent fluctuations and smaller correlation length-scales as it approaches perihelia, consistent with the expectation of ``younger'' turbulence close to the Sun \citep[e.g.,][]{bruno2013LRSP}. A mix of Alfv\'enic and low cross-helicity states are observed, with the latter concentrated around the heliospheric current sheet region. Increasing solar activity (via increasing dipole tilt) leads to larger variation in the levels of measured turbulence quantities. 

We also test  the Taylor ``frozen-in'' hypothesis along the planned mission trajectory, finding low levels of validity for the standard approximation near perihelia. A number of modified ``frozen-in'' approximations are also tested, yielding generally unsatisfactory results. The expected failure of the Taylor hypothesis implies that the space turbulence community must seek alternative frameworks for the interpretation of primary mission observations.

A variety of other relevant quantities may be calculated from the turbulence parameters presented here, such as energetic charged particle diffusion coefficients, 
turbulent heating rates, rates of Coulomb collisions, 
model turbulence spectra, and so on, potentially enabling a 
variety of other studies relevant to the Probe mission. 
To this end, the data presented in the paper will be made available as Supplementary Material online. 

As a final remark we note that the present approach, once updated with closer to real-time magnetograms, 
is expected to become useful for planning observation strategy during the mission, and later for retrospective data interpretation.
\section{Acknowledgments}
We thank J. Kasper for useful discussions and the Johns Hopkins University Applied Physics Laboratory's  \href{https://sppgway.jhuapl.edu/}{\psp~project office}
for providing the \href{https://naif.jpl.nasa.gov/naif/index.html}{NASA \textit{SPICE} kernel} containing the \psp~ephemeris. This research is supported in part by the NASA 
\textit{Parker Solar Probe} mission
through the IS\(\sun\)IS project and subcontract SUB0000165
from Princeton University to University of Delaware, by 
the NASA Heliophysics Grand Challenge program under grant NNX14AI63G, the NASA Living With a Star program under grant NNX15AB88G, the NASA Heliospheric Guest Investigator program through grant NNX17AB79G, and by NASA Heliospheric Supporting Research grants 80NSSC18K1210 and 80NSSC18K1648.
%
\bibliographystyle{aasjournal}

%

\end{document}